\documentclass[prl,amsmath,amssymb,twocolumn,longbibliography]{revtex4-1}
\usepackage{graphicx}
\usepackage{dcolumn}
\usepackage{bm}
\usepackage{placeins}
\usepackage{rotating}
\usepackage{multirow}
\usepackage{longtable}
\usepackage[dvipsnames,usenames]{xcolor}
\usepackage[english]{babel}
\usepackage{amssymb} 
\usepackage{amsmath} 
\usepackage[utf8]{inputenc} 
\usepackage[normalem]{ulem}

\newcommand{\be}{\begin{equation}}
\newcommand{\ee}{\end{equation}}
\newcommand{\bea}{\begin{eqnarray}}
\newcommand{\eea}{\end{eqnarray}}

\newcommand{\Tr}{ {\rm Tr} \, }

\def\a{\alpha}

\def\G{\Gamma}

\def\l{\lambda}

\def\S{\Sigma}

\def\xc{{\rm xc}}
\def\x{{\rm x}}

\def\Tr{{\rm Tr}\,}

\begin{document}
\widetext 

\title{High-pressure II-III phase transition in solid hydrogen: Insights \\from state-of-the-art ab initio calculations}
\author{Maria Hellgren}
\affiliation{Sorbonne Universit\'e, MNHN, UMR CNRS 7590, IMPMC, 4 place Jussieu, 75005 Paris, France}
\author{Damian Contant}
\affiliation{Sorbonne Universit\'e, MNHN, UMR CNRS 7590, IMPMC, 4 place Jussieu, 75005 Paris, France}
\author{Thomas Pitts}
\affiliation{Sorbonne Universit\'e, MNHN, UMR CNRS 7590, IMPMC, 4 place Jussieu, 75005 Paris, France}
\author{Michele Casula}
\affiliation{Sorbonne Universit\'e, MNHN, UMR CNRS 7590, IMPMC, 4 place Jussieu, 75005 Paris, France}
\date{\today}
\begin{abstract}
The high-pressure II-III phase transition in solid hydrogen is investigated using the random phase approximation and diffusion Monte Carlo. 
Good agreement between the methods is found confirming that an accurate treatment of exchange and correlation increases the transition pressure by more than 100 GPa with respect to semilocal density functional approximations. Using an optimized hybrid functional, we then reveal a low-symmetry structure for phase II generated by an out-of-plane librational instability of the C2/c phase III structure. 
This instability weakens the in-plane polarization of C2/c leading to the well-known experimental signatures of the II-III phase transition such as a sharp shift in vibron frequency, infrared activity and $c/a$ lattice parameter ratio. Finally, we discuss the zero-point vibrational energy that plays an important role in stabilizing phase III at lower pressures.
\end{abstract}
\maketitle

The phase diagram of pure hydrogen has intrigued and challenged theoretical and experimental physicists for decades. Despite being the simplest element, high-pressure hydrogen forms complex solid phases governed by strongly interacting electrons and quantum nuclei \cite{rev2,mao1994ultrahigh,rev1}. 
Present knowledge suggests that at low pressure hydrogen consists of freely rotating molecules centered on a hcp structure (phase I). At 110 GPa the Raman roton bands disappear \cite{PhysRevLett.64.1939,PhysRevLett.80.101} and small vibron discontinuities are observed \cite{PhysRevLett.70.3760}. This is the onset of the broken symmetry phase (phase II), in which the rotational motion is hindered due to the increasing anisotropic intermolecular interactions, while maintaining strong fluctuating behavior due to the relevance of nuclear quantum effects (NQEs)\cite{biermann1998quantum,geneste2012strong,PhysRevB.84.144119,PhysRevB.55.11330}. Around 150 GPa hydrogen enters phase III. The II-III phase transition has been the subject of several studies \cite{hemley1994charge,mazin1995insulator,cui1995excitations,mazin1997quantum,tosatti,edwards2004order,toledano2009symmetry,jcp2011}. The transition is detected by the sharp drop of vibron frequency  \cite{PhysRevLett.61.857,PhysRevLett.63.2080}, combined with a rapid increase in infrared (IR) activity \cite{PhysRevLett.70.3760}. The spectral signatures are experimentally well-established and the structures are not expected to largely deviate from the hcp symmetry \cite{PhysRevB.82.060101}. However, little is still known concerning the orientational order. Although several structures have been proposed based on theoretical considerations \cite{PhysRev.167.862,PhysRevLett.68.2468,natoli1995crystal,kohanoff1997solid,tosatti,johnson2000structure,TSE20085,doi:10.1063/1.3679749}, large uncertainties remain since their relative energies have been difficult to accurately determine with first-principles calculations. This is mainly due to the approximate treatment of the electron-electron interaction \cite{PhysRevB.89.184106,drummond2015quantum,liao2019comparative}, but also due to difficulties in including
 NQEs \cite{MonserratSCF,Azadi2017shissor,Rillo_2018,SCHA2021,Morresi2021}. 

 The most promising candidates for phase III are layered structures, in particular one of C2/c symmetry  \cite{pickard2007structure,Eremets2019_hydro,Loubeyre2019observation,Gorelov2020,MonacelliNatPhys2020}, 
and a more recent one of P6$_1$22 symmetry \cite{PhysRevB.94.134101,PhysRevB.100.155103}.
The planar arrangement of the H$_2$ molecular units induces a polarization and stretches the H$_2$ bond length. This leads to the strong IR activity and the softened vibron frequency when compared to the many phase II candidate structures, which all contain
canted molecules with respect to the hcp planes \cite{tosatti,doi:10.1063/1.3679749}. Most approximations within density functional theory (DFT) predict the static (i.e. clamped nuclei) II-III transition to occur below or around the experimental value at about 150-155 GPa \cite{jcp2011,PhysRevB.82.060101}. However, more accurate diffusion Monte Carlo (DMC) \cite{drummond2015quantum} and coupled cluster single double (CCSD) \cite{liao2019comparative} calculations have shown to shift the static transition pressure beyond 250 GPa. Adding zero-point vibrational energies from DFT to the DMC enthalpies reduces this result by 20 GPa only \cite{drummond2015quantum}. 
These results either questions the structures or suggests that the relative role played by the electron-electron interaction and lattice dynamical effects is not well understood.
 
In this work, we reexamine the II-III phase transition in solid hydrogen using state-of-the-art \emph{ab initio} calculations based on the random phase approximation (RPA) and DMC. 
We discover a new low-symmetry structure for phase II that is
stabilized by nuclear vibrations
and, unlike previous candidate structures, emerges from
a continuous symmetry breaking of phase III.
This rationalizes several experimental outcomes. 
Furthermore, we show that the transition pressure is compatible with experimental findings when accurate electronic energies are
 considered, together with
zero-point energy variations.
  We finally provide an
  estimate of NQEs, in terms of quantum anharmonicity across the transition.

Many previous works have applied DMC to hydrogen, and it provides the gold standard for this system \cite{PhysRevB.89.184106,drummond2015quantum,McMinis_2015,monacelli2022quantum}. 
Here, we use DMC in its lattice regularized version \cite{casula2005diffusion} to project an initial variational wave function of Jastrow-Slater form, with Slater orbitals generated by DFT within the local density approximation. Further details can be found in the supplemental material (SM) \cite{SM}.
\begin{figure*}[t!]
\includegraphics[scale=0.45,angle=0]{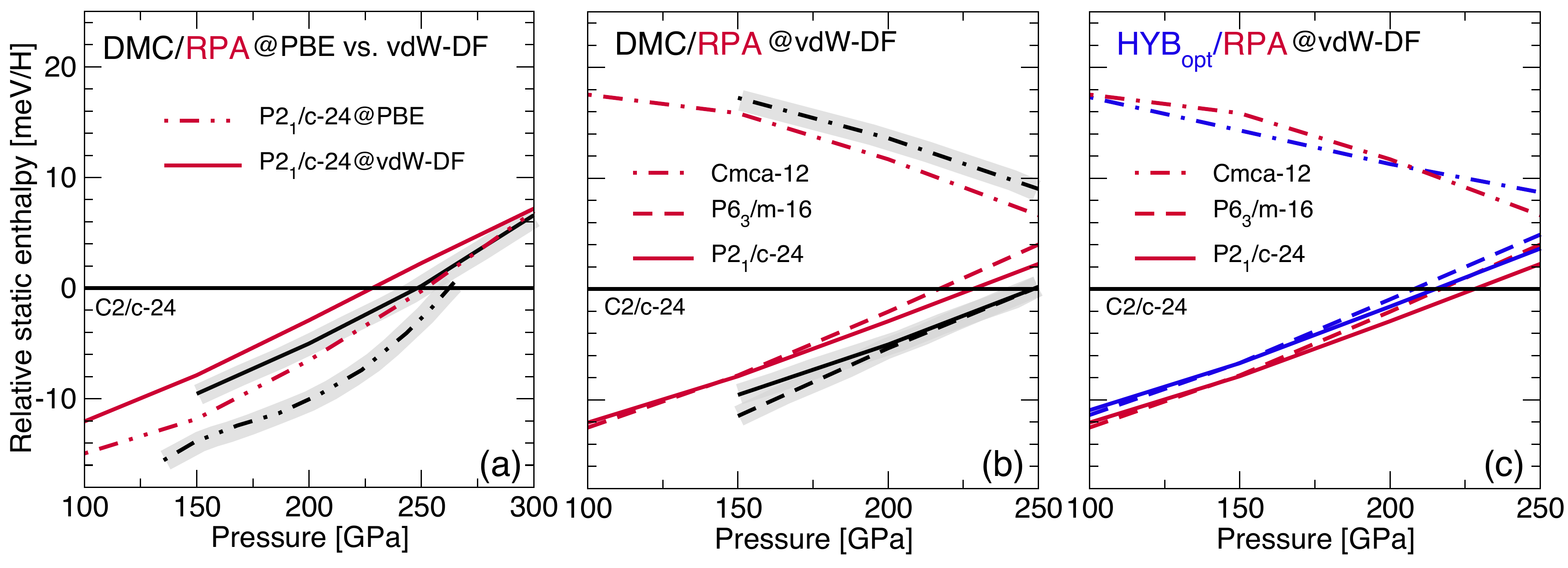}
\caption{RPA (red) compared to DMC (black, (a) and (b)) and HYB$_{\rm opt}$ (blue, (c)). The @DFT notation refers to the DFT approximation used to optimise the geometry at fixed pressure. The P2$_1$/c-24@PBE result is extracted from Ref. {\cite{drummond2015quantum}}. The DMC error bars are indicated by a gray shaded area, having a height of $\pm 0.9$ meV, around the mean values.}
\label{rpadmc}
\end{figure*}

The RPA is known for high accuracy at moderate computational cost \cite{ren2012random,PhysRevB.79.205114,PhysRevLett.103.056401}, and is here applied to hydrogen for the first time. 
It combines exact exchange with a formally exact expression for the correlation energy $E_{\rm c}$ written in terms of the dynamical linear density response function $\chi_\l$
\be
E_{\rm c}= -\int_0^1d\lambda\int_0^\infty\frac{d\omega}{2\pi}\Tr \{v[\chi_\l(i\omega)-\chi_s(i\omega)]\}.
\label{lambda_integration}
\ee
Within the RPA $\chi_\l(i\omega)$ fulfills the time-dependent Hartree equations: $\chi_\l(i\omega)=\chi_s(i\omega)+\chi_s(i\omega) \l v\chi_\l(i\omega)$, where $\chi_s(i\omega)$ is the independent-particle Kohn-Sham response function and $v$ is the Coulomb interaction. Including a vertex via the exact-exchange kernel leads to the RPA with exchange (RPAx), which has proven to give more reliable energy differences due to systematically improved total energies \cite{doi:10.1063/1.3290947,doi:10.1080/00268970903476662,PhysRevB.90.125150,doi:10.1063/1.4922517,PhysRevB.98.045117,PhysRevResearch.3.033263}. Here, we will use the RPAx, not only as an additional validation of the RPA, but also to optimize the fraction of exchange in an approximate hybrid functional.

We start by demonstrating the performance of RPA in the 100-300 GPa pressure range. We study several structures previously proposed in the literature (denoted by their symmetry and number of atoms); P2$_1$/c-24, P6$_3$/m-16 and Pca2$_1$-8 for phase II and C2/c-24 and Cmca-12 for phase III. Within a given symmetry the geometry is optimized using the vdW-DF functional \cite{PhysRevLett.92.246401}. Previous calculations have shown that this functional gives accurate geometries at fixed volume for molecular solid hydrogen  \cite{PhysRevB.89.184106}.  In Fig.~\ref{rpadmc}(a) we report the results for the static enthalpy difference between P2$_1$/c-24 and C2/c-24 with DMC and RPA, as well as the result from an earlier DMC calculation that used structures optimized with PBE \cite{drummond2015quantum}. For comparison, RPA results obtained on PBE structures are also presented \footnote{All calculations are done with the Quantum ESPRESSO package \cite{qe} using an ONCV (Optimized Norm-Conservinng Vanderbilt) pseudopotential \cite{Hamann_2013}. More details can be found in the SM.}. First of all, we see that the shift in transition pressure due to the change of functional used to optimize the geometry is of the order 20 GPa with both DMC and RPA. Secondly, RPA is found to be in very good agreement with DMC, staying consistently within 2~meV/H around the DMC mean value over the whole pressure range.

In panel \ref{rpadmc}(b), P6$_3$/m-16 and Cmca-12 are also included. A recent CCSD calculation of static enthalpies predicted the phase II candidate P6$_3$/m-16 to be the most stable phase up to 350 GPa \cite{liao2019comparative}. Our DMC calculation contradicts this result, with P6$_3$/m-16 being degenerate with P2$_1$/c-24 within error bars ($\pm 0.9$ meV). This behaviour is also confirmed by our RPA calculation, and it is independent of the theory used to optimize the geometry. There is also a good agreement between DMC and RPA for the Cmca-12 
structure. Cmca-12 was originally a candidate for phase III but is now expected to become important at higher pressures, close to the insulator-to-metal transition \cite{Loubeyre_2002,McMinis_2015,Eremets2019_hydro,Loubeyre2019observation,Gorelov2020,MonacelliNatPhys2020,monacelli2022quantum}. 

We have also carried out RPAx calculations on the same structures.
Including exchange in the response function has a very small effect on the RPA energy differences in these systems, as shown in the 
SM \cite{SM}. Thus, the agreement between RPA, RPAx and DMC provides strong confirmation that the  static II-III transition pressure, with the currently known structures, should lie at about 225-250 GPa, irrespective
of whether P2$_1$/c-24 or P6$_3$/m-16 is used for phase II. We will later show that, according to our most accurate calculations, a third competing symmetry for phase II, i.e. Pca2$_1$-8, lies very close in energy to the other two phases.
These results imply an overestimation of 70-100 GPa with respect to experiment. 
 This is also independent of whether C2/c-24 or P6$_1$22-36 is used for phase III.
  Indeed, they can be
  considered degenerate within the DMC error bars in the pressure range analyzed here (see SM). However, we chose C2/c-24 as a reference structure for phase III in what follows.
\begin{figure}[t]
\includegraphics[scale=0.43,angle=0]{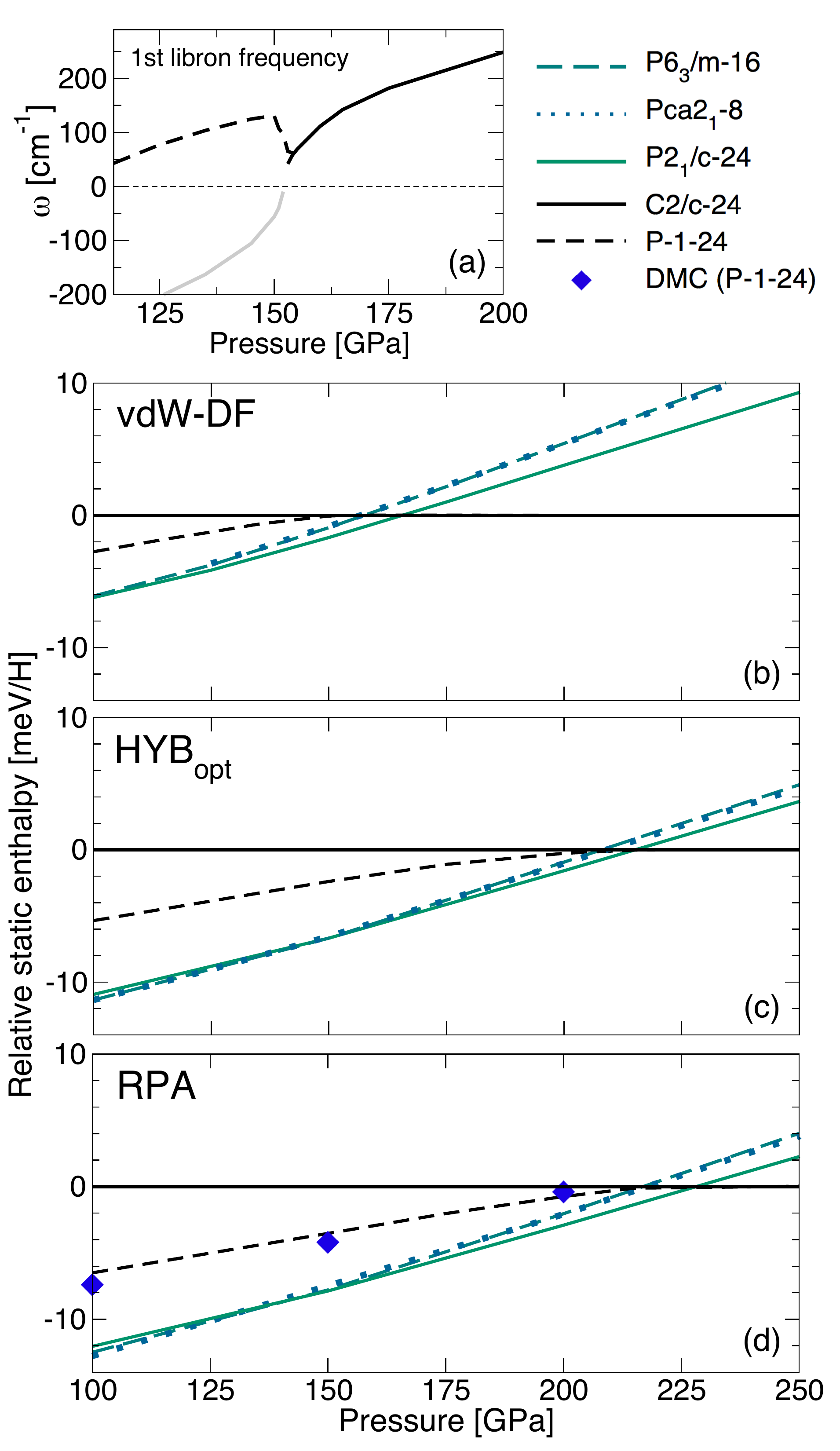}
\caption{The frequency of the unstable vdW-DF libron as a function of pressure is shown in (a), with the imaginary frequency (grey) shown as a negative frequency. Relative static enthalpy of P2$_1$/c-24, P6$_3$/m-16 and Pca2$_1$-8 with respect to C2/c-24 for (b) vdW-DF, (c) HYB$_{\rm opt}$ and (d) RPA. The enthalpy of the structure obtained by following the librational instability of C2/c-24 is also displayed and denoted P-1-24. The DMC results for P-1-24 are marked with blue diamonds in panel (d). Their size spans the range covered by the DMC error bars.}
\label{enthalpy}
\end{figure}

To investigate the reason for the difference with respect to experiment and to include effects of lattice vibrations, we now analyse the possibility of using a hybrid functional that retains the QMC and RPA accuracy at a cheaper cost. The standard PBE0 functional with 25\% of exchange does not improve the enthalpy differences with respect to semilocal DFT functionals (see SM \cite{SM}). However, exact-exchange is clearly of crucial importance in the molecular phases of solid hydrogen since PBE and Hartree-Fock alone give II-III transition pressures at 110 GPa and beyond 450 GPa, respectively \cite{liao2019comparative}. We will, therefore, optimize a new fraction of exchange using the accurate RPAx total energy. The optimization is carried out by minimizing the RPAx total energy of
an isolated
H$_2$ molecule with respect to the fraction of exchange used to generate the input density \cite{tise2_prb,oepdg}. The approach is described in Ref. \cite{PhysRevResearch.3.033263} and in the SM \cite{SM}. We find a minimum at 48\%, which is well beyond the standard value. The results from this optimized hybrid functional, which we denote as HYB$_{\rm opt}$, are presented in Fig.~\ref{rpadmc} (c). The good agreement with RPA shows that a hybrid functional with a carefully chosen exact-exchange fraction is sufficient to produce accurate enthalpies. 
\begin{figure}[b]
\includegraphics[scale=0.31,angle=0]{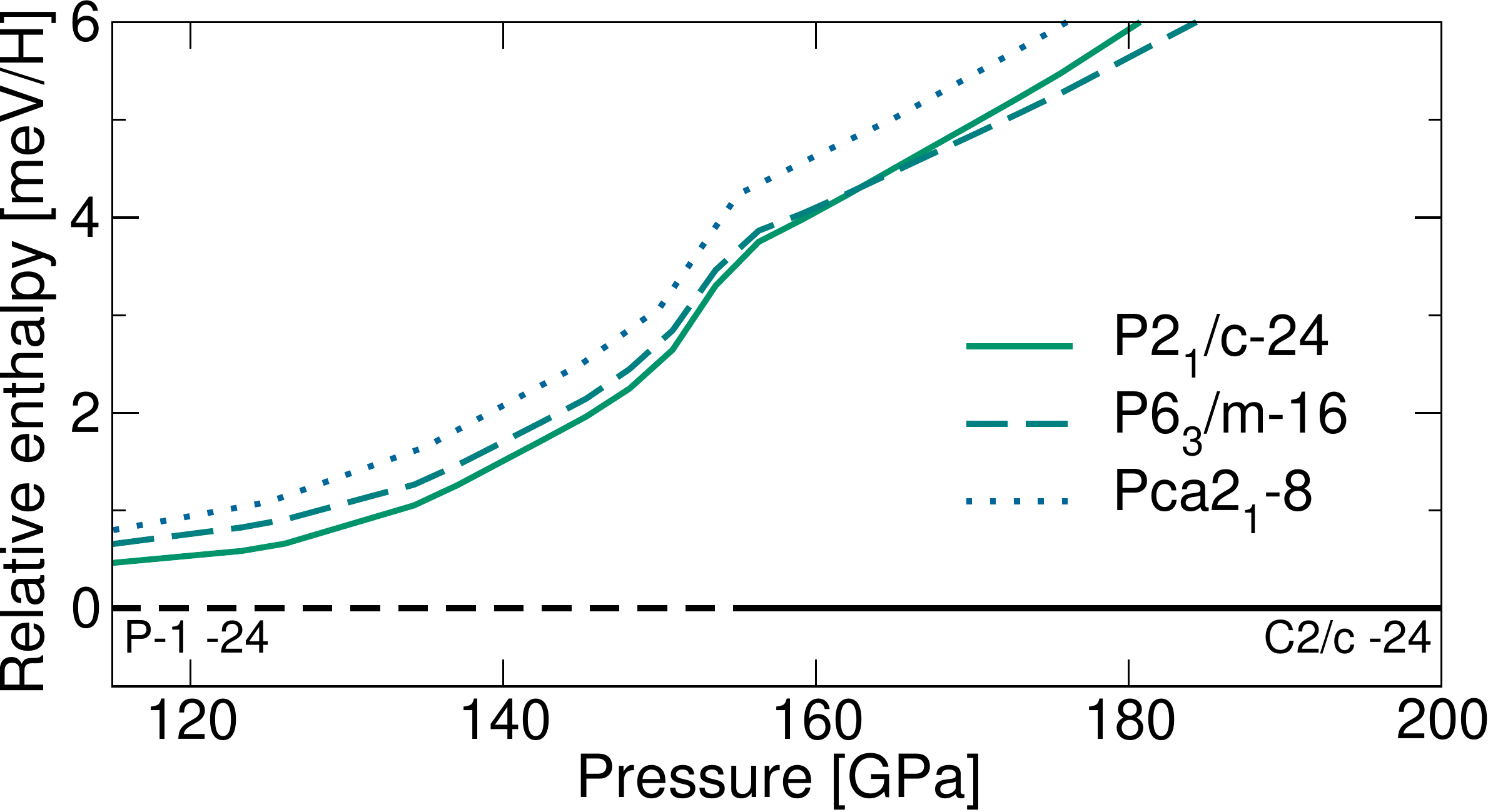}
\caption{Relative enthalpies within vdW-DF including ZPEs in the harmonic approximation. The ZPE of P-1-24 is smaller than the ZPE of P2$_1$/c-24, P6$_3$/m-16 and Pca2$_1$-8, reversing their order of stability.}
\label{zpe}
\end{figure}

 We can now use HYB$_{\rm opt}$ to study the stability of the structures and the impact of their relaxed geometry \footnote{Structural relaxation with hybrid functionals was performed using the VASP code \cite{vasp1,vasp2,vasp3}.}. Comparing enthalpy differences using vdW-DF and HYB$_{\rm opt}$ geometries gives a difference of less than 1 meV/H. The static enthalpy differences that we have obtained with vdW-DF geometries are thus robust to further variations in the geometry. The stability of the structures can then be studied by calculating the vibrational spectra. With HYB$_{\rm opt}$ we are limited to ${\bf \G}$-point vibrations, but with vdW-DF we can study dense ${\bf q}$-point grids. We found P2$_1$/c-24 and P6$_3$/m-16 to both be stable with vdW-DF in the pressure range investigated. Interestingly, C2/c-24 is found to have ${\bf \G}$-instabilities (two nearly degenerate imaginary phonons) below 215 GPa with HYB$_{\rm opt}$ and below 150 GPa with vdW-DF, i.e. exactly at the expected static II-III transition within the given functional. The lowest vdW-DF libron that becomes imaginary is shown as a function of pressure in Fig. 2 (a). We note that a similar behaviour has been observed by Raman spectroscopy \cite{hemley1990low,mazin1997quantum,PhysRevLett.80.101,goncharov2001spectroscopic}, suggesting the onset of a libron instability within phase III, in proximity to the transition pressure. The unstable libron mode generates a new structure with lower symmetry, in which two thirds of the C2/c-24 molecules are rotated out of plane (the structures are visualized in the SM \cite{SM}). The space-group symmetry is thereby reduced to P-1, i.e., only inversion symmetry remains. By displacing the atoms in C2/c-24 according to the phonon mode eigenvector, we find a symmetric double-well potential. However, the minimum of this potential is strongly underestimated since the molecular tilting should be accompanied by an H$_2$ bond length contraction and an expansion of the $c/a$ lattice parameter ratio. Indeed, after a full geometry relaxation, the energy gain increases by two orders of magnitude. Using our HYB$_{\rm opt}$ optimized geometries, the lowering of energy is confirmed at both the RPA and DMC level (Fig.~\ref{enthalpy}(d)). 

In Fig.~\ref{enthalpy} we summarize the results for the phase diagram with clamped nuclei, including also the stable Pca2$_1$-8 phase II structure.
Comparing DMC, RPA, HYB$_{\rm opt}$ and vdW-DF a clear trend emerges. 
The position of the instability in C2/c-24 coincides with the transition between C2/c-24 and all the proposed phase II structures, which are all nearly degenerate, independently of which functional is used. Furthermore, all approximations predict the P-1-24 structure to lie above the phase II candidates. Therefore, P-1-24 does not appear competitive at any pressure. However, this picture changes when we consider lattice vibrations.  

Let us now include lattice dynamics via the harmonic zero-point vibrational energy (ZPE). Due to the presence of the C2/c-24 instability, it is necessary to use the same functional for the electronic energy as for generating the structures and computing the ZPEs. Previous calculations used PBE structures, where such an instability occurs below the range of interest (at 110 GPa) \cite{drummond2015quantum}. Calculating the ZPE with RPA and HYB$_{\rm opt}$ is presently not feasible due to the high computational cost of using supercells. However, we can consistently include the ZPE within the cheaper vdW-DF functional \footnote{Zero-point energies are calculated using density functional perturbation theory as implemented within the QE PHonon code \cite{RevModPhys.73.515,qe}.}. The result can be found in Fig.~\ref{zpe}. We immediately see that the corresponding picture is different. The phase II candidates are all shifted by roughly the same amount, falling above the C2/c-24 and P-1-24 structures. Below 150 GPa, P-1-24 is now the most stable. Given the similarity in qualitative behaviour between the different functionals, it is plausible that we would find the same ordering of enthalpies with the more advanced electronic structure methods.
\begin{figure}[t]
\includegraphics[scale=0.36,angle=0]{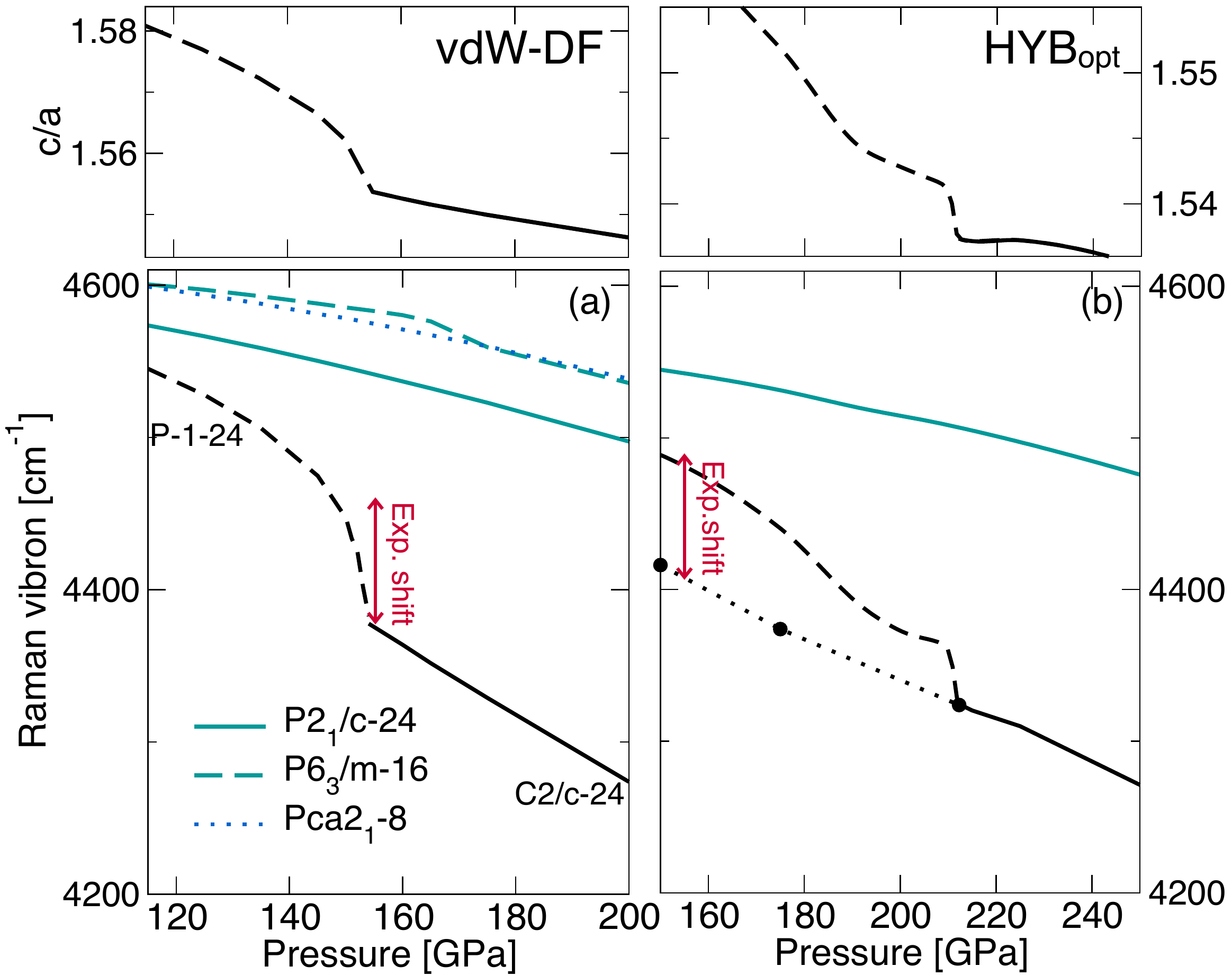}
\caption{(a) The lowest vibron frequency as a function of pressure within vdW-DF for P2$_1$/c-24, P6$_3$/m-16, Pca2$_1$-8 and C2/c-24. A rapid upshift in the frequency is observed at the transition from C2/c-24 (full line) to P-1-24 (dashed line). A similar shift is seen in the $c/a$ ratio, as shown in the upper panel. (b) The same results with HYB$_{\rm opt}$ for P2$_1$/c-24 and C2/c-24. Black dots joined by the dotted line correspond to the frequencies obtained from the unstable C2/c-24 structure. The experimental shift of 80 cm$^{-1}$ at 155 GPa \cite{PhysRevLett.63.2080} is marked in both panels (vertically upshifted with respect to experiment).}
\label{vibron}
\end{figure}

Having shown that P-1-24 is a promising candidate for phase II, let us move to an in-depth analysis of the transition.
A well-established experimental feature of the II-III phase transition is the sharp shift of 80 cm$^{-1}$ in vibron frequency \cite{PhysRevLett.61.857,PhysRevLett.63.2080,Loubeyre_2002,PhysRevB.82.060101}. In Fig.~\ref{vibron} we plot the lowest vibron frequency, which is Raman active, as a function of pressure. The left panel (a) shows the vdW-DF results for C2/c-24, P2$_1$/c-24, P6$_3$/m-16 and Pca2$_1$-8 and the right panel (b) shows the HYB$_{\rm opt}$ results for C2/c-24 and P2$_1$/c-24. The change of C2/c-24 into P-1-24 is marked by dashed lines. At 155 GPa, i.e. at the transition to P-1-24, the C2/c-24 vdW-DF vibron exhibits a continuous but sharp increase, whose size is very similar to the experimental result marked in red \cite{PhysRevLett.63.2080}. There is an overall difference of approximately 400 cm$^{-1}$ with respect to experiment that can only be accounted for when calculating the vibrations beyond the harmonic approximation \cite{MonacelliNatPhys2020,morresi2022hydrogen}. We note that the frequency shift going from C2/c-24 to P-1-24 agrees better with experiment than going from C2/c-24 to any of the other candidate phase II structures. We expect effects of anharmonicity to mostly cancel but any effect would most likely increase the shift since the C2/c-24 vibron is more anharmonic due to the stronger in-plane interactions. The HYB$_{\rm opt}$ result is qualitatively similar to the vdW-DF result, although the vibron shift appears sharper and is smaller than with vdW-DF. However, when extrapolated to 155 GPa using the unstable C2/c-24 structure (i.e. when calculated at the experimental transition pressure) the shift again agrees well with experiment. 

So far, C2/c-24 has only been challenged by the similar P6$_1$22-36 structure \cite{PhysRevB.94.134101,PhysRevB.100.155103}. 
To show that the mechanism we found is common to these phase-III type of structures, we have repeated the calculations above for P6$_1$22-36. Indeed, the vibron shift is almost identical and caused by a librational instability very similar to the one in C2/c-24. The results can be found in the SM \cite{SM}. One could reasonably assume that other energetically competitive planar structures, if found for example by structural searches using a functional beyond PBE, are likely to exhibit the same feature.

The cause of the abrupt change in vibron frequency is related to the shortening of the intramolecular bond lengths in P-1-24. Indeed, the out-of-plane rotation of the H$_2$ units weakens the in-plane intermolecular interactions that stretch the bond length. At the same time, the $c$-parameter of the nearly hexagonal lattice abruptly increases. In the upper panels of Fig.~\ref{vibron} (a) and (b) we have plotted the $c/a$ lattice parameter ratio as a function of pressure. We observe a shift very similar to the
one in vibron frequency \cite{kitamura2000quantum}. This behaviour has also been observed by X-ray diffraction 
\cite{PhysRevB.82.060101}.
In addition, by studying the charge distribution, we find that the polarization of some of the molecules reduces by a factor of two when rotated out of plane (see SM \cite{SM} for a Bader analysis \cite{bader,badercode} of the charges in C2/c-24 and P-1-24). This change is reflected in the IR activity, which rapidly decreases moving from phase III to phase II \cite{PhysRevLett.70.3760}. In the SM we present the IR intensity as a function of pressure using the vdW-DF functional. A qualitatively good agreement with experiment is found.

Let us finally discuss the vibrational contribution to the P-1-24 to C2/c-24 transition pressure. The exact calculation of the transition pressure, including all effects of vibrations, represents a very difficult task. Indeed, the electron-electron interaction should be described at least at the level of a hybrid functional and lattice vibrations should be calculated beyond the harmonic approximation. However, we can make an initial estimate by calculating the variation in the harmonic ZPE due to the change in vibrons only. The 12 vibrons are only weakly ${\bf k}$-dependent so we can make the estimate at the ${\bf \G}$-point. We find that the P-1-24 energy increases by around 2~meV/H with both vdW-DF and HYB$_{\rm opt}$. This would already lower the transition pressure by 40 GPa. 
Moreover, quantum anharmonicity strongly affects the orientational symmetry breaking of the molecular in-plane order. Indeed, the instability driven by the lowest librational modes can be modelled by a double well potential, as previously mentioned. According to this simple model, derived from HYB$_{\rm opt}$ energies, NQEs reduce the transition pressure by an additional amount of 20 GPa, due to quantum resymmetrization effects (see SM) \cite{Nature2016}. A very mild isotope effect is found, in accordance with experiments\cite{goncharov1995invariant,edwards2004order}. This brings the transition pressure obtained with the most advanced electronic structure methods to a value much closer to experiment, corroborating the mechanism of the transition.

In conclusion, using a combination of RPA, DMC, hybrid DFT and vdW-DF functionals, we have provided new insights into the nature of the II-III phase transition. We have revealed the existence of a libron instability in C2/c-24 that generates a new broken symmetry phase when the pressure is lowered, in which two thirds of the H$_2$ molecules are rotated out of plane. This relatively small orientational change is sufficient to quantitatively reproduce the experimental signatures of a sharp vibron shift and an order of magnitude increase of IR intensity, at the pressure where the system undergoes the transition into phase III.

\begin{acknowledgements}
The work was performed using HPC resources from GENCI-TGCC/CINES/IDRIS (Grants No. A0110907625 and A0110906493). 
Financial support from Emergence-Ville de Paris is acknowledged.
This  work  was partially supported  by  the  European  Centre of Excellence in Exascale Computing TREX-Targeting Real Chemical Accuracy at the Exascale, funded by the European Union's Horizon 2020 Research and Innovation program under Grant Agreement No.~952165.
\end{acknowledgements}
%
\begin{widetext}
\section{Supplemental materials for: \\
High-pressure II-III phase transition in solid hydrogen: Insights from state-of-the-art ab initio calculations}
Below we present additional information about the computational
details, structural visualization, a complementary analysis of the P6$_1$22-36 structure,
the nuclear quantum effects analysis across the II-III phase
transition, Bader charge analysis and infrared
spectra.

\section{QMC calculations details}

Diffusion Monte Carlo (DMC) calculations have been performed using the lattice regularized DMC scheme (LRDMC) \cite{casula2005diffusion}, as implemented in the TurboRVB package \cite{nakano2020turborvb}.
DMC allows one to access properties of the quantum many-body distribution that is obtained by projecting the initial variational wave function toward the ground state of the system within the fixed-node (FN) approximation.

To initialize the many-body state, we employed a Jastrow-Slater variational wave function
$\Psi^\textbf{k}(\textbf{R})=\exp\{-U(\textbf{R})\} \det\{\phi^\textbf{k}_j(\textbf{r}^\uparrow_i)\}\det\{\phi^\textbf{k}_j(\textbf{r}^\downarrow_i)\}$ for $i,j \in \{1,\ldots,N/2\}$, where $N$ is the number of electrons in the unpolarized supercell, $\textbf{k}$ is the twist belonging to a Monkhorst-Pack (MP) grid of the supercell Brillouin zone, and $\textbf{R}=\{\textbf{r}^\uparrow_1,\ldots,\textbf{r}^\uparrow_{N/2},\textbf{r}^\downarrow_1,\ldots,\textbf{r}^\downarrow_{N/2}\}$ is the $N$-electron coordinate. 

$U$ is the Jastrow function, split into three contributions:  $U=U_{en}+U_{ee}+U_{een}$.
The electron-nucleus function $U_{en}$ has an exponential decay and is given by $U_{en}=\sum_{iI} J_{1b}(r_{iI}) + U_{en}^\textrm{no-cusp}$, where the index $i$($I$) runs over electrons (nucleus), $r_{iI}$ is the electron-nucleus distance, and $J_{1b}(r)=\alpha (1-\exp\{-r/\alpha\})$, with a variational parameter $\alpha$ . $J_{1b}$ cures the nuclear cusp conditions, and allows the use of the bare Coulomb potential in our QMC framework. The electron-electron function $U_{ee}$ has a Pad\'e form and is given by $U_{ee}=-\sum_{i \ne j} J_{2b}(r_{ij})$, where the indices $i$ and $j$ run over electrons, $r_{ij}$ is the electron-electron distance, and $J_{2b}(r)=0.5 r/(1+ \beta r)$, with a variational parameter $\beta$. This two-body Jastrow term fulfills the cusp conditions for antiparallel electrons. The last term in the Jastrow factor is the electron-electron-nucleus function: $U_{een}=\sum_{(i \ne j) I} \sum_{\gamma\delta} M_{\gamma \delta I}\chi_{\gamma I}(r_{iI}) \chi_{\delta I}(r_{jI})$, with $M_{\gamma \delta I}$ a matrix of variational parameters, and $\chi_{\gamma I}(r)$ a $(2s,2p,1d)$ Gaussian basis set, with orbital index $\gamma$, centered on the nucleus $I$. 
Analogously, the electron-nucleus cusp-free contribution to the Jastrow function, $U_{en}^\textrm{no-cusp}$, is developed on the same Gaussian basis set, such that  $U_{en}^\textrm{no-cusp}=\sum_{i I} \sum_{\gamma} V_{\gamma I}\chi_{\gamma I}(r_{iI})$, where $V_{\gamma I}$ is a vector of parameters.
The $J_{1b}$ and $J_{2b}$ Jastrow functions are made periodic using a $\textbf{r} \rightarrow \textbf{r}^\prime$ mapping that makes the distances diverge at the border of the unit cell, as explained in Ref.~\cite{nakano2020turborvb}. For the inhomogeneous $U_{een}$ part and for the electron-nucleus term $U_{en}^\textrm{no-cusp}$, the Gaussian basis set $\chi$ is made periodic by summing over replicas translated by lattice vectors.

The one-body orbitals $\phi$ are expanded in a primitive $(4s,2p,1d)$ Gaussian basis set, which we contracted into 6 hybrid orbitals using the geminal embedding orbitals (GEO) contraction scheme \cite{sorella2015geminal} at the $\Gamma$ point. The $\phi$ orbitals are made periodic using the same scheme as for the $\chi$ orbitals. This basis set yields a FN-LRDMC bias in the energy differences smaller than the target error of 1 meV per atom. 
For each $\textbf{k}$ belonging to the MP grid of a given supercell (see Tab.~\ref{tab:kmesh}), we performed independent DFT calculations in the local density approximation (LDA) to generate $\{\phi^\textbf{k}_j\}_{j=1,\ldots,N/2}$ for all occupied states. These LDA calculations are done for the fully \emph{ab initio} Hamiltonian with bare Coulomb potential for the electron-ion interactions. The same Hamiltonian is then solved using LRDMC.

Before running LRDMC calculations, we optimized the $\alpha$, $\beta$,
$M_{\gamma \delta I}$ and $V_{\gamma I}$ parameters, keeping the
orbitals $\phi^\textbf{k}_i$ fixed. These parameters are optimized by
minimizing the variational energy of the wave function $\Psi$ within
the QMC linear optimization method \cite{umrigar2007alleviation}. All
$\textbf{k}$-twists share the same Jastrow factor. The LRDMC
projection is carried out at the lattice space $a=0.25 a_0$, where
$a_0$ is the Bohr radius, yielding
converged energy differences.
\begin{figure}[htb!]
\includegraphics[scale=0.35]{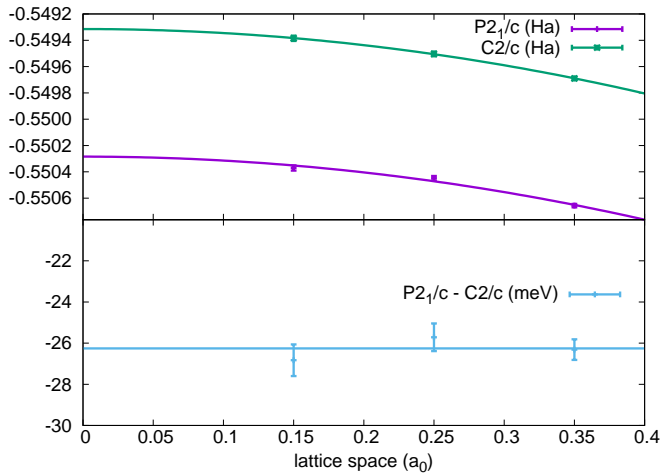}
\caption{Lattice space convergence of
 LRDMC total energies for the P2$_1$/c and C2/c structures at a volume
 corresponding to 150 GPa. Upper panel: total energies extrapolations
 in Hartree/H. Lower panel: constant extrapolation of the energy
 difference in meV/H. In energy differences,  convergence is reached already at $a=0.35 a_0$.}
\label{fig:a_convergence}
\end{figure}
For example, in Fig.~\ref{fig:a_convergence} we show the lattice space
extrapolation of the total energy for the C2/c and the P2$_1$/c
symmetries, computed at 150 GPa, as well as the lattice space
extrapolation of their energy difference. As one can see, energy
differences are converged in lattice space already for larger values
of $a$ than the one used in this work.
This is an advantage with respect to the ``standard'' diffusion Monte
Carlo algorithm, which sometimes has a less smooth convergence in the
time step. Indeed, in LRDMC the lattice space extrapolation has an $a^2$
behavior, which is given by the lattice regularization of the kinetic
term. 

For LRDMC equilibration, we discarded the first
5 blocks from the averages of each twist, to get rid of the transient
regime. A single block is made of 25 branching steps, each step is
made of 100 moves on the LRDMC random lattice. This corresponds to an
imaginary time evolution of 0.43 Ha$^{-1}$ per branching step. The time
propagation for each twist is carried by 32 walkers, within a fixed
walker population algorithm. This is enough for a stable projection,
given the quality of our starting variational wave function. The finite size
population bias has been removed by the ``correcting factors
scheme'' \cite{buonaura1998numerical}. This sampling strategy allows us to reach a target statistical
precision of about 1 meV/H, once the statistics is gathered from all
$\bf{k}$-twists used.


\begin{table}[h!]
    \centering
    \begin{tabular}{c | c | c | c }
         crystal symmetry & $\textbf{k}$-mesh size & supercell size  & $N$ \\
         \hline
         C2/c-24 & $6\times 12 \times 6$ & $2 \times 1 \times 2$ & 96 \\
         P-1-24 &  $6\times 12 \times 6$ & $2 \times 1 \times 2$ & 96 \\
         Cmca-12 & $6\times 6\times 6$ & $2 \times 2 \times 2$ & 96\\
         P2$_1$/c-24 & $12\times 6\times 6$ & $1 \times 2 \times 2$ & 96 \\
         P6$_3$/m-16 & $6 \times 6 \times 6$ & $2 \times 2 \times 2$ & 128 \\
    \end{tabular}
    \caption{\textbf{k}-meshes and supercell sizes employed in our QMC
      simulations of crystal symmetries shown in the first column. The
      last column reports the number of atoms $N$ in the corresponding
      supercell. \textbf{k}-meshes are expressed with respect to
      the supercell Brillouin zone. } 
    \label{tab:kmesh}
\end{table}

 In our QMC calculations, we used $\textbf{k}$-meshes and
 supercell sizes reported in Tab.~\ref{tab:kmesh}. To further reduce
 finite-size errors, we used Kwee-Zhang-Krakauer (KZK) corrected
 energies \cite{kwee2008finite}. The KZK corrections as a function of
 pressure and crystal symmetry are reported in Tab.~\ref{tab:KZK} for the smallest supercell size taken
      into account in the present work. As one can see, the KZK
      correction dependence on the specific crystal symmetry at fixed
      pressure and fixed size is negligible within 1 meV/H
      accuracy. Therefore, KZK corrections largely cancel out in
      the energy differences at the given pressure, if the supercell
      size is the same.
\begin{table}[h!]
    \centering
    \begin{tabular}{c | c | c | c | c }
         crystal symmetry & 150 GPa & 200 GPa  & 250 GPa & 300 GPa \\
         \hline
         C2/c-24 ($N$=96) & 91.7  & 97.5  & 102.0  & 105.9 \\
      P2$_1$/c-24  ($N$=96) & 91.5 & 97.0 & 101.6 & 105.5 \\
      Cmca-12 ($N$=96) & 92.0 & 97.6 & 102.2 & 106.1 \\
      P6$_3$/m-16 ($N$=128) & 68.4 & 72.6 & 76.0 & 78.9 \\
    \end{tabular}
    \caption{Magnitude of KZK corrections for the pressures analyzed,
      and for different structures at the smallest supercell size taken
      into account in the present work (reported in parenthesis). The numerical entries are in meV/H.
    }
    \label{tab:KZK}
\end{table}

We ran all LRDMC calculations long enough to reach a statistical error
of about 1 meV per atom ($\pm 0.9$ meV/H for $N$=96). 
 For the systems and pressures studied here,
 this computational setup leads to systematic
 finite-size errors whose magnitude falls into the statistical error bar.
 
To check for the presence of a finite-size bias, we performed
additional calculations with a larger supercell, $N$=288, for C2/c,P2$_1$/c
and P6$_3$/m at pressures of 150 GPa and 200 GPa, and compare with the
results obtained from smaller systems. The results are
reported in Tab.~\ref{tab:FS_errors}, where ``regular supercell'' is the one with
at least 96 atoms, used throughout the paper, while ``large supercell''
is the one with 288 atoms.
\begin{table}[h!]
    \centering
    \begin{tabular}{c | c | c | c | c }
         systems &  supercell size & $\Delta$E (meV/H)  &
                                                          $\sigma_{\Delta\textrm{E}}$
                                                          (meV/H) &
                                                                   large/regular
                                                                   variation
      \\
      & & & & in units of $\sigma$\\
      \hline
      \hline
      P2$_1$/c - C2/c (150 GPa) & regular  & -9.5  & 0.7  &  \\
      P2$_1$/c - C2/c (150 GPa) & large  & -8.2  & 1.2  & 1.0 \\
      \hline
      P2$_1$/c - C2/c (200 GPa) & regular &  -5.0 & 0.7 &  \\
      P2$_1$/c - C2/c (200 GPa) & large &  -7.1 &  1.2 & -1.5 \\
      \hline
      P6$_3$/m - C2/c (150 GPa) & regular & -11.4 & 0.9 &  \\
      P6$_3$/m - C2/c (150 GPa) & large & -7.8 & 1.3 & 2.3 \\
      \hline
      P6$_3$/m - C2/c (200 GPa) & regular & -5.3 & 0.8 &  \\
      P6$_3$/m - C2/c (200 GPa) & large & -4.2 & 1.3 & 0.7 \\
    \end{tabular}
    \caption{Energy difference fluctuations due to the change of
      supercell size. The variation is reported in statistical error
      ($\sigma$) units in the last column.
    }
    \label{tab:FS_errors}
\end{table}
As it is apparent from the Table, the $\Delta$E variation between the two sizes falls within 2
standard deviations in absolute value. Thus, the residual FS bias, if present, is of the order of the
statistical error bar.
 
To compute enthalpies from internal energies, we employed the
pressures estimated at the HYB$_\textrm{opt}$ level. We checked that
these pressures are in statistical agreement with the QMC pressures,
which have however an error bar of $\approx 2$ GPa on
average. Therefore, to keep the error bar on the final enthalpies
small, we used the HYB$_\textrm{opt}$ pressures in the transformation.

\section{Details of the DFT and RPA calculations}
Density functional calculations based on PBE and vdW-DF have been performed with Quantum ESPRESSO (QE) \cite{qe} and an ONCV (Optimized Norm-Conserving Vanderbilt) pseudopotential \cite{Hamann_2013}. For structural optimization we used 110 Ry plane-wave cutoff and up to $8\times 8\times 8$ ${\bf k}$-point grids. Zero point energies were calculated using density functional perturbation theory as implemented within the PHonon package. Hybrid functional calculations were performed with QE on vdW-DF and PBE geometries. The volume was optimized by minimizing the hybrid enthalpy at fixed pressure. The relative enthalpies of P2$_1$/c-24 and P6$_3$/m-16 with respect to C2/c-24 on PBE geometries can be found in Fig. ~\ref{fig:convergence}, including also the CCSD results of Ref. \cite{liao2019comparative} and the DMC results obtained in this work.

\begin{figure}[htb!]
\includegraphics[scale=0.45]{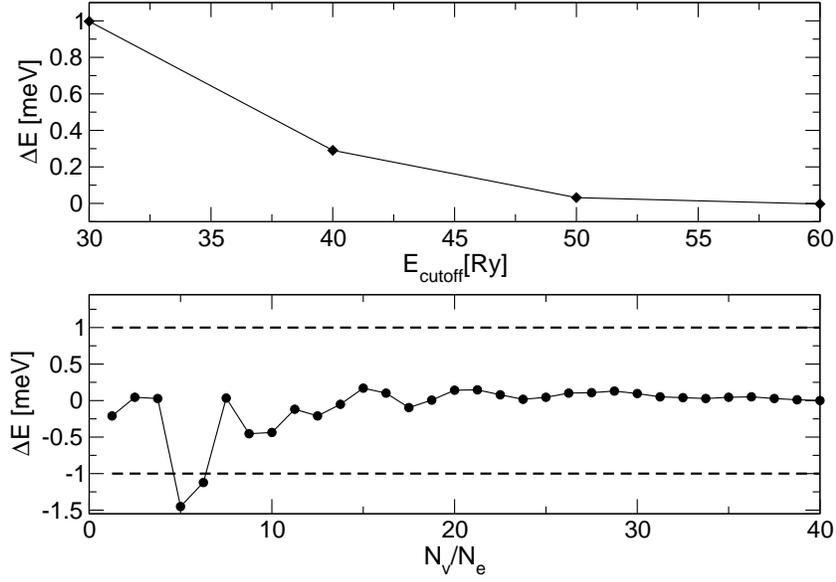}
\caption{Convergence of the plane-wave cutoff energy (top) and the
  number of eigenvalues used in the response function (bottom) in the
  RPA calculations.}
\label{fig:convergence}
\end{figure}

Full structural optimization using hybrid functionals was carried out with the VASP code \cite{vasp1,vasp2,vasp3} using the PAW method and 400 eV plane-wave cutoff. The charge density for the Bader analysis \cite{bader,badercode} was also obtained with the VASP code using the vdW-DF functional. 

The RPA and RPAx calculations were performed using an implementation based on an eigenvalue decomposition of the response function \cite{PhysRevB.79.205114,PhysRevB.90.125150,PhysRevResearch.3.033263}. For the correlation energies we used a plane-wave cutoff energy of 60 Ry and 8 eigenvalues per electron for the response function. The convergence with respect to these two parameters can be found in Fig. 3. The ${\bf k}$-points were converged at $5\times 3\times 5$ (C2/c-24), $2\times 7\times 2$ (P2$_1$/c-24), $5\times 5\times 6$ (P6$_3$/m-16), $5\times 8\times 6$ (Pca2$_1$-8) and $4\times 4\times 4$ (Cmca-12). In RPAx we used smaller grids to save computational cost [$4\times 3\times 4$ (C2/c-24), $3\times 4\times 3$ (P2$_1$/c-24), $4\times 4\times 4$ (P6$_3$/m-16), and $4\times 4\times 4$ (Cmca-12)] but staying within 2 meV of accuracy. In Fig.~\ref{fig:rpax_vs_rpa} we present a comparison between RPA and RPAx. 

The $\alpha$-parameter of the hybrid functional is optimized by numerically minimizing the total energy in RPAx with respect to the $\alpha$ used to generate the input density, i.e.
\be
\frac{d E^{\rm RPAx}[n^{\alpha}]}{d {\alpha}}=0.
\ee
The local xc potential of the hybrid functional needed to generate $n^{\alpha}$ is defined as the functional derivative of the hybrid xc energy with respect to the density. It can thus be decomposed into \cite{tise2_prb}
\be
v_\xc^{\rm hyb,\a}=v^{\a}_{\x}+(1-\a)v^{\rm PBE}_{\x}+v^{\rm PBE}_{\rm c}.
\label{lhyb}
\ee
For the exchange part, an integral equation known as the linearized Sham-Schl\"uter equation has to be solved
\be
\int\!d2\, \chi_s(1,2) v^{\a}_{\x}(2)=-i\a\int\!d2d3\,G_s(3,1)G_s(1,2) \S_s^{\rm HF}(2,3)
\label{exxlss}
\ee
where $\S_s^{\rm HF}$ is the Fock self-energy, $G_s$ is the Kohn-Sham
Green's function and $\chi_s$ is the Kohn-Sham independent particle
response function. This equation is solved using an iterative
technique as described in Ref.~\cite{oepdg}. The result for an
isolated H$_2$ molecule is presented in Fig.~\ref{fig:opt_alpha}. 

The RPA/RPAx energies where evaluated on top of PBE orbitals. We have verified that 
using orbitals from a local hybrid potential with an optimal fraction of exchange does not change the energy differences significantly.

\begin{figure}[t]
\begin{center}
\includegraphics[scale=0.35]{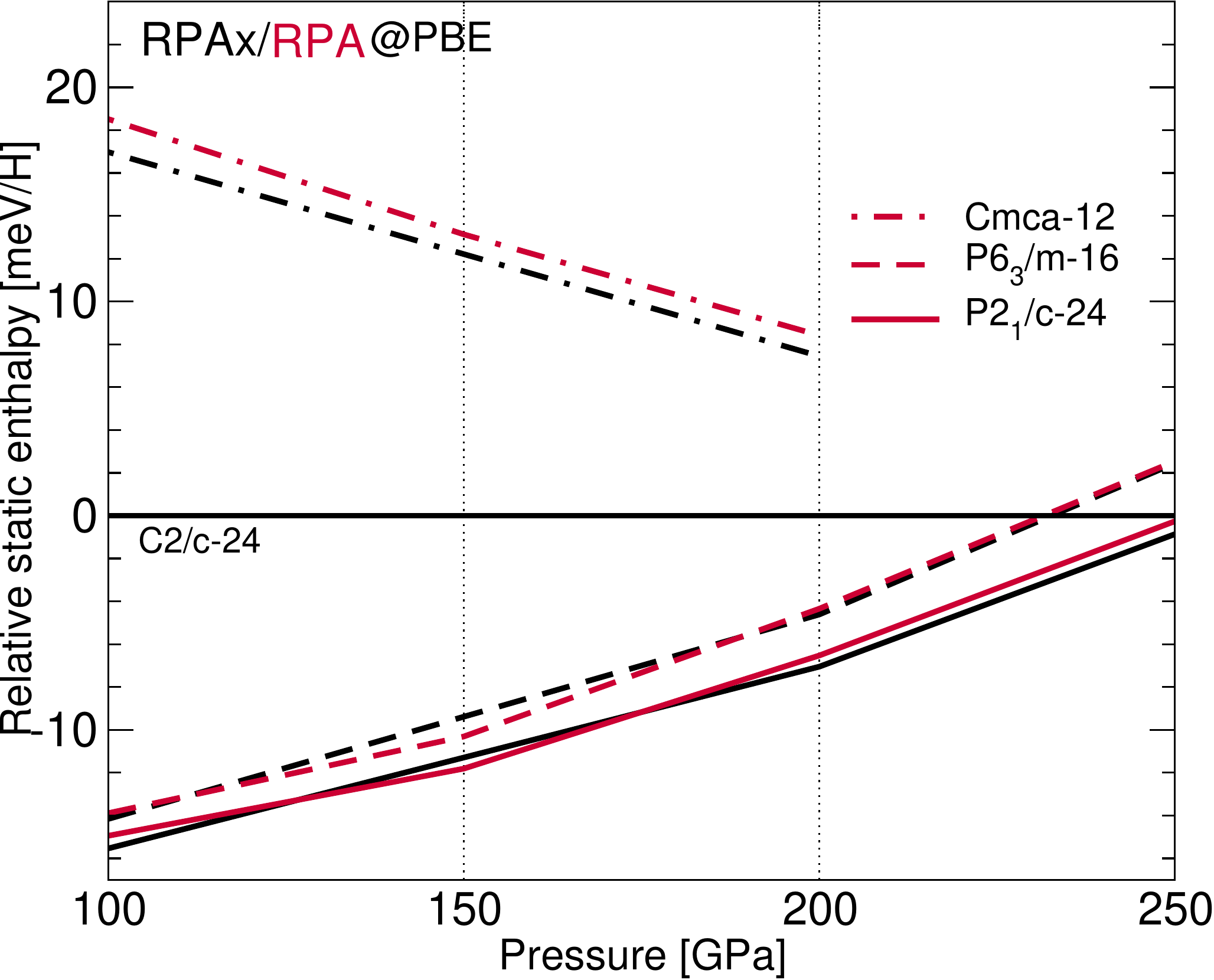}
\end{center}
\caption{RPAx compared to RPA.}
\label{fig:rpax_vs_rpa}
\end{figure}

\begin{figure}[t]
\begin{center}
\includegraphics[scale=0.35]{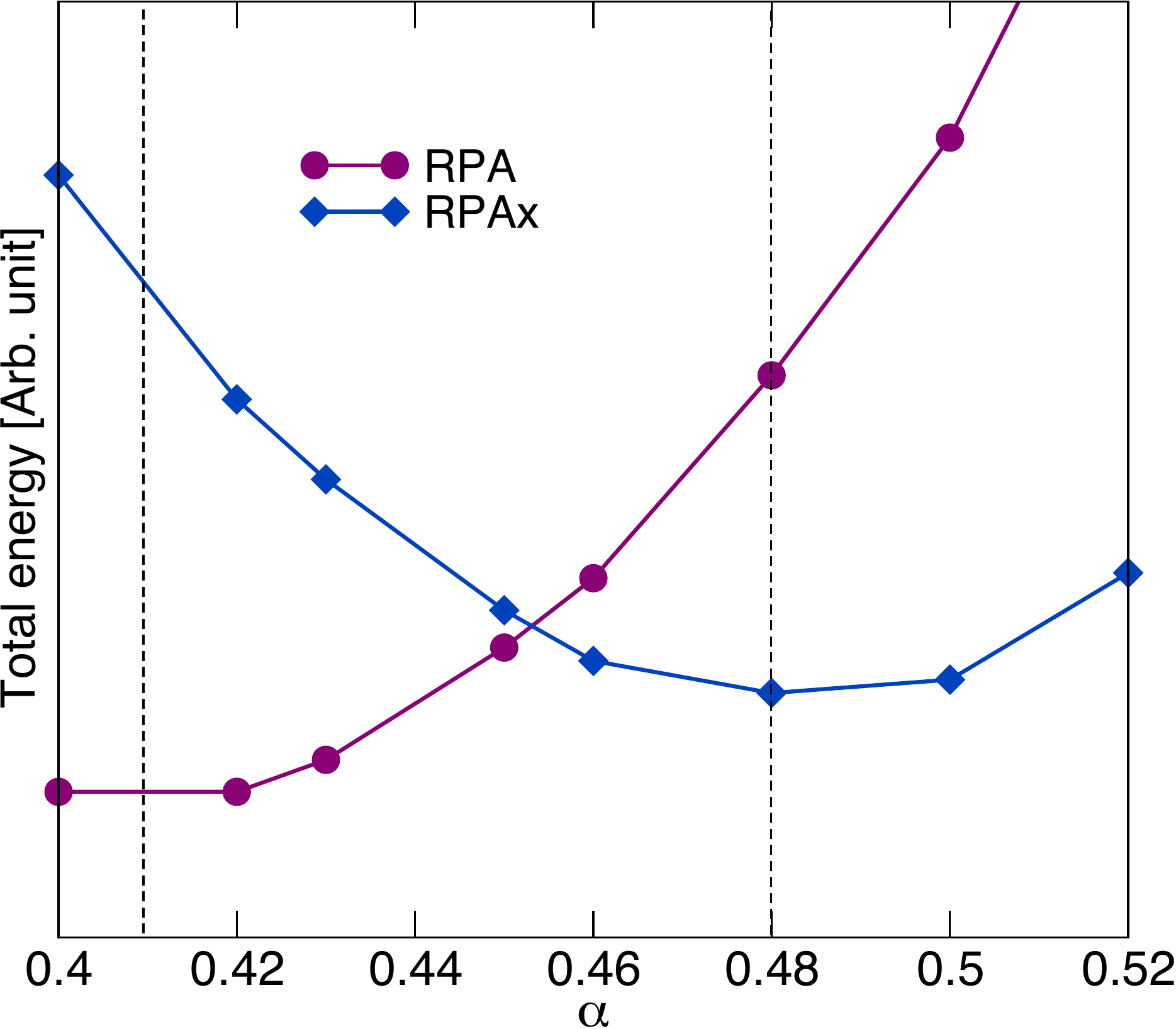}
\end{center}
\caption{The $\a$-parameter of the hybrid functional is determined by
  locating the minimum of the RPAx total energy on the H$_2$
  molecule.}
\label{fig:opt_alpha}
\end{figure}


\section{The II-III transition pressure with various approximations to 
  the exchange and correlation energy}

\begin{figure}[h!]
\includegraphics[scale=0.45,angle=0]{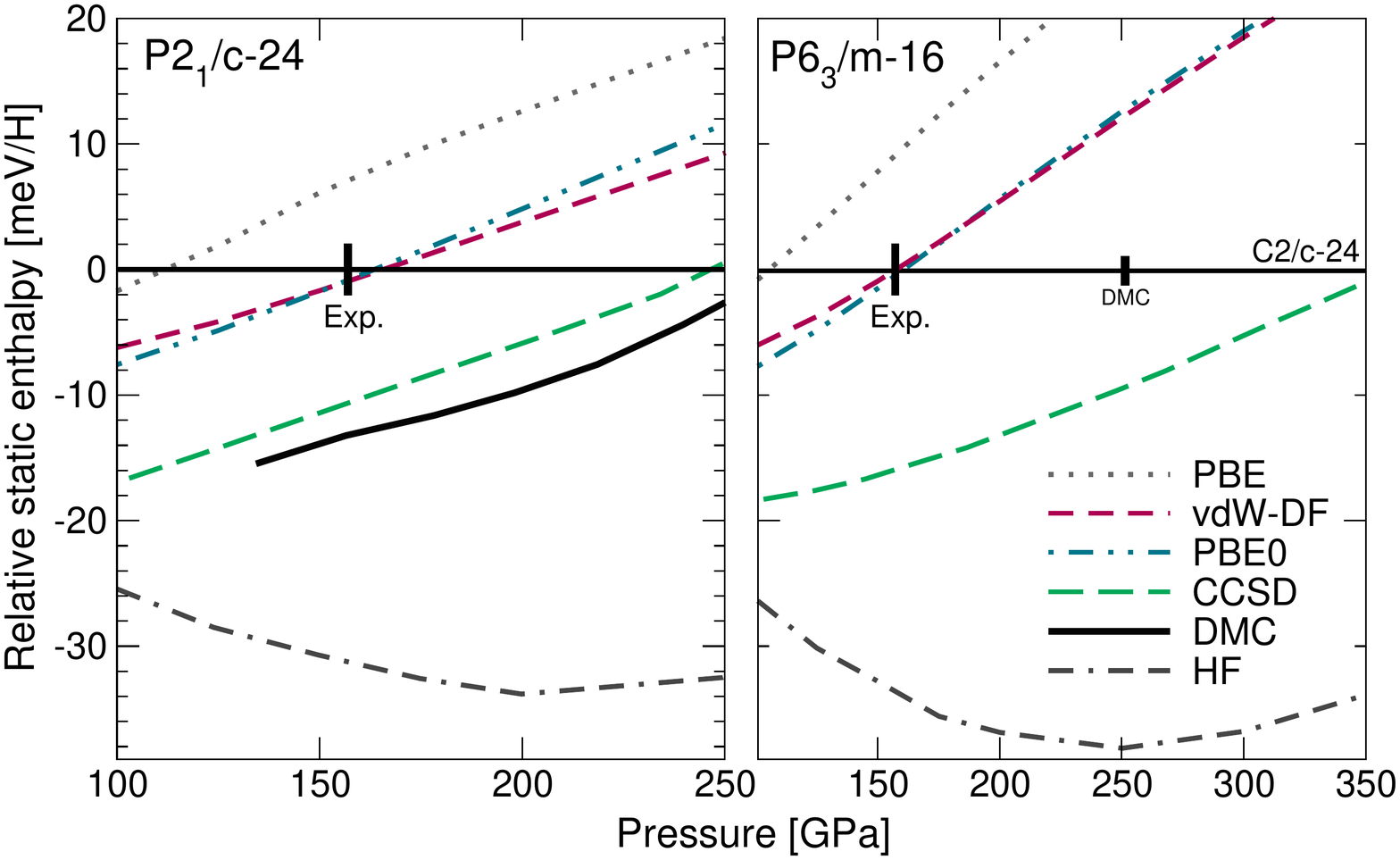}
\caption{The II-III transition pressure with various approximations to 
  the exchange and correlation energy (PBE, vdW-DF, PBE0, CCSD, DMC 
  and HF).}
\label{fig:compare_pbe}
\end{figure}

\newpage

\section{The C2/c-24 and P-1-24 structures}

\begin{figure}[h!]
\includegraphics[scale=0.4]{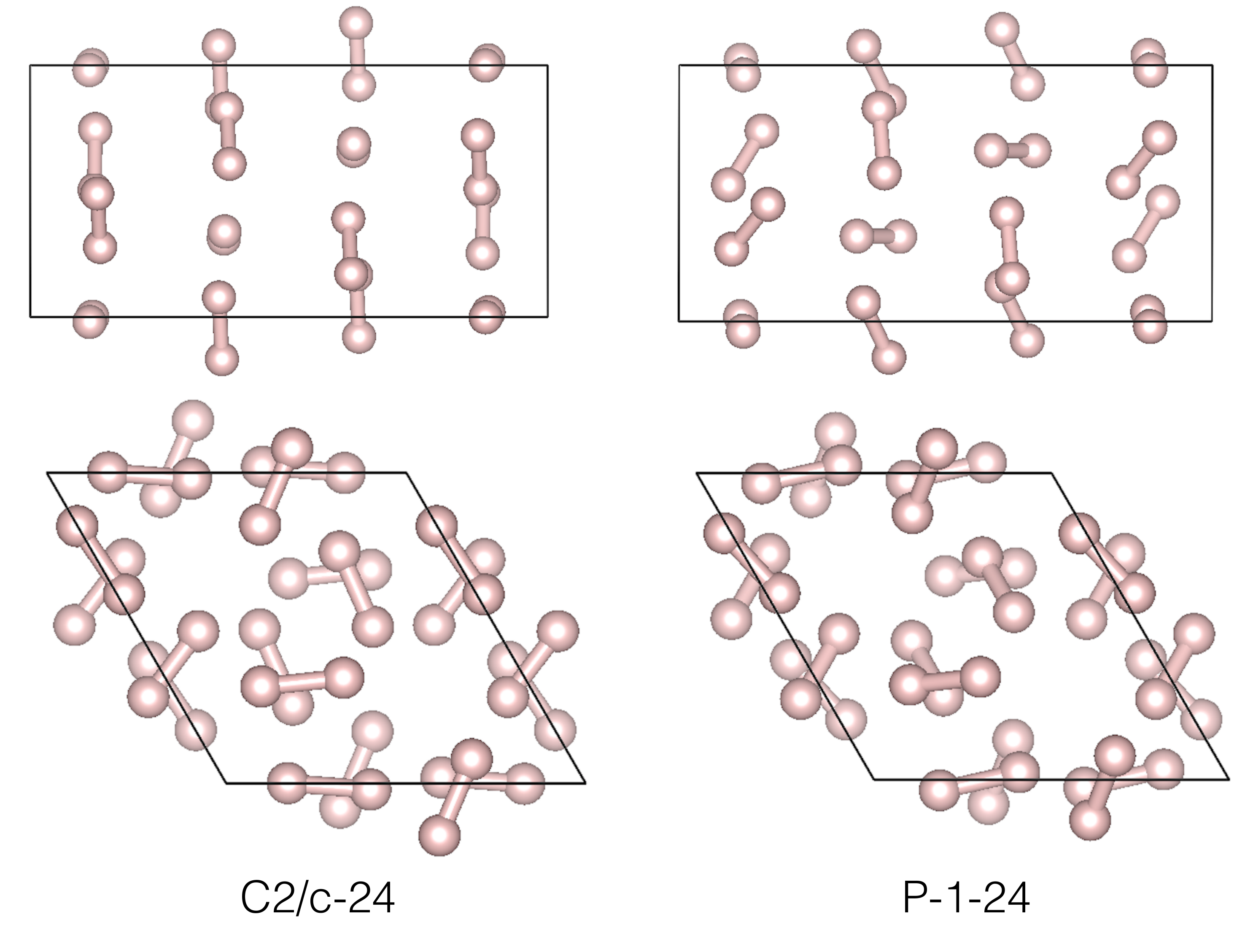}
\caption{View along the c-axis (top) and the a-axis (bottom) of
  C2/c-24 (left) and P-1-24 (right).}
\label{fig:structures}
\end{figure}

\section{Analysis of the P6$_1$22-36 symmetry}

With the aim at quantifying the relative stability between the
P6$_1$22-36 structure and the C2/c-24 one in the 150-250 GPa pressure
range, we carried out additional extensive QMC calculations at 150
GPa. According to Ref.~\cite{PhysRevB.100.155103}, at this pressure the energy
difference between these two phases should be maximized and in favor
of P6$_1$22 by at least 17 meV/H. We performed LRDMC simulations up to a system size of 432
atoms, with $\bf{k}$-points sampling (corresponding to a
$12\times 12\times 12$
Monkhorst-Pack grid for the unit cell) and KZK 2-body corrections, and
we found that the C2/c and P6$_1$22 are nearly degenerate in enthalpy,
with a small 2.5 $\pm$
1.1 meV gain for the C2/c structure. With the reached QMC statistical
accuracy, they can be considered as degenerate within two error
bars. The finite-size extrapolation is plotted in Fig.~\ref{fig:DMC_P6122} for
both phases, showing the DMC total enthalpies for the sizes we computed
in order to resolve their energy difference in the thermodynamic
limit. The largest evaluated size, N=432, corresponds to a 2x2x3
supercell for the P6$_1$22 crystal symmetry.
\begin{figure}[htb!]
\begin{center}
\includegraphics[scale=0.35]{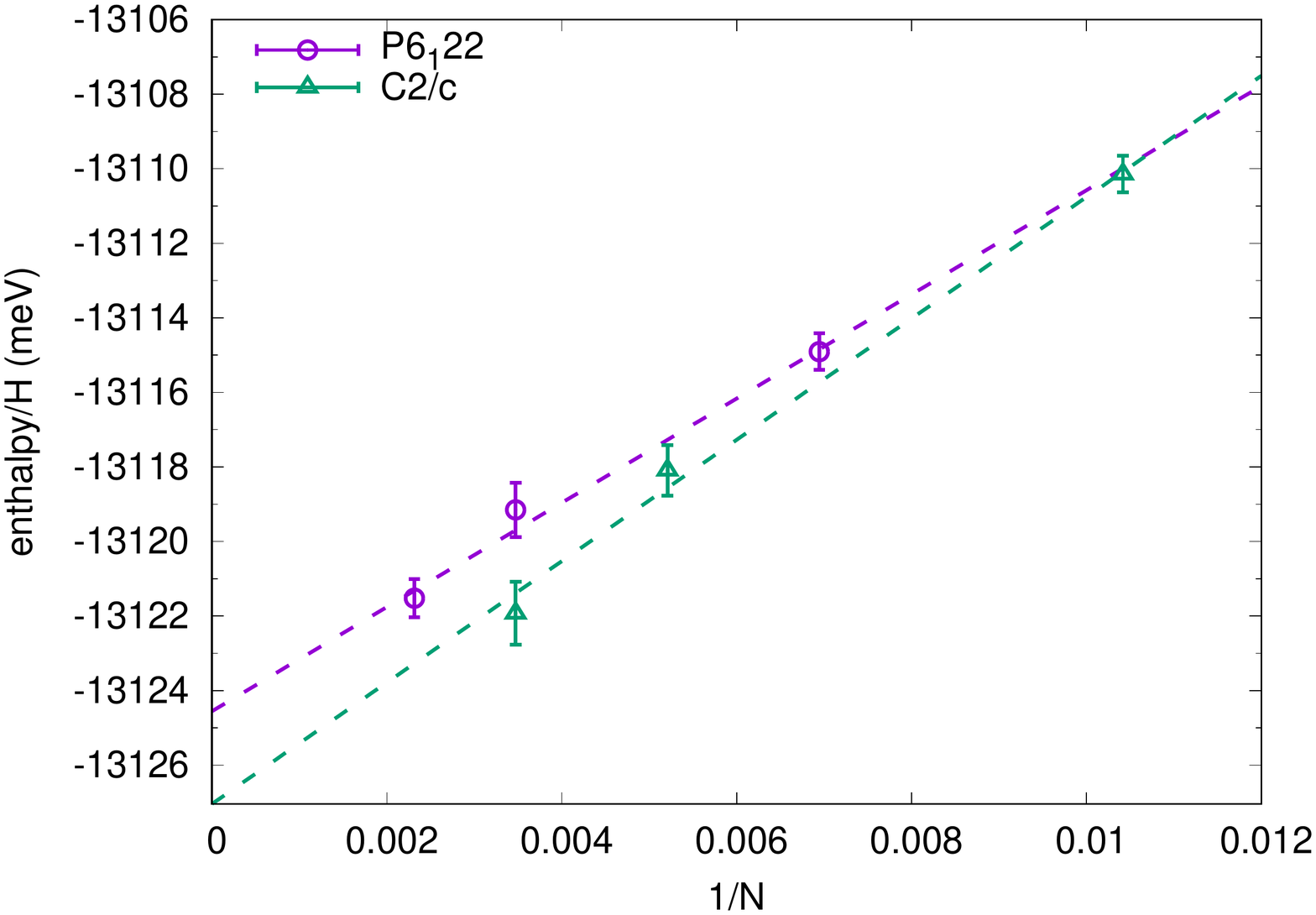}
\end{center}
\caption{Total enthalpies of P6$_1$22 and C2/c structures, for different system sizes. 
For C2/c, we considered the 2x1x2, 2x2x2, and 2x3x2 supercells, while
for P6122 we computed the 2x2x1, 2x2x2 and 2x2x3 supercells. The
largest one computed has N=432 hydrogen atoms. The dashed lines
indicate the extrapolated enthalpies.} 
\label{fig:DMC_P6122}
\end{figure}
The above result is in agreement with HYB$_\textrm{opt}$ calculations that we
performed in addition to our QMC results. This outcome is at variance
with Ref.~\cite{PhysRevB.100.155103}, which instead reports the
P6$_1$22 structure as the most stable one.
We attribute the source of this discrepancy to a too coarse
$\bf{k}$-mesh employed in that work. Nevertheless, the two structures
stay very close to each other in enthalpy, and a possible polymorphism of phase
III in this pressure range, favored by nuclear quantum effects (NQEs), is likely.

Since the P6$_1$22 structure is very similar to C2/c, one would expect
similar structural instabilities that can explain the II-III
transition. As shown in Fig.~\ref{fig:p6122_transition},  we have
verified that the P6$_1$22 has a libron instability similar to the one
of C2/c, giving rise to a vibron jump with the same magnitude. The
only difference with respect to C2/c is
that it occurs at a slightly larger pressure (by 5 GPa).
 
\begin{figure}
\includegraphics[scale=0.45]{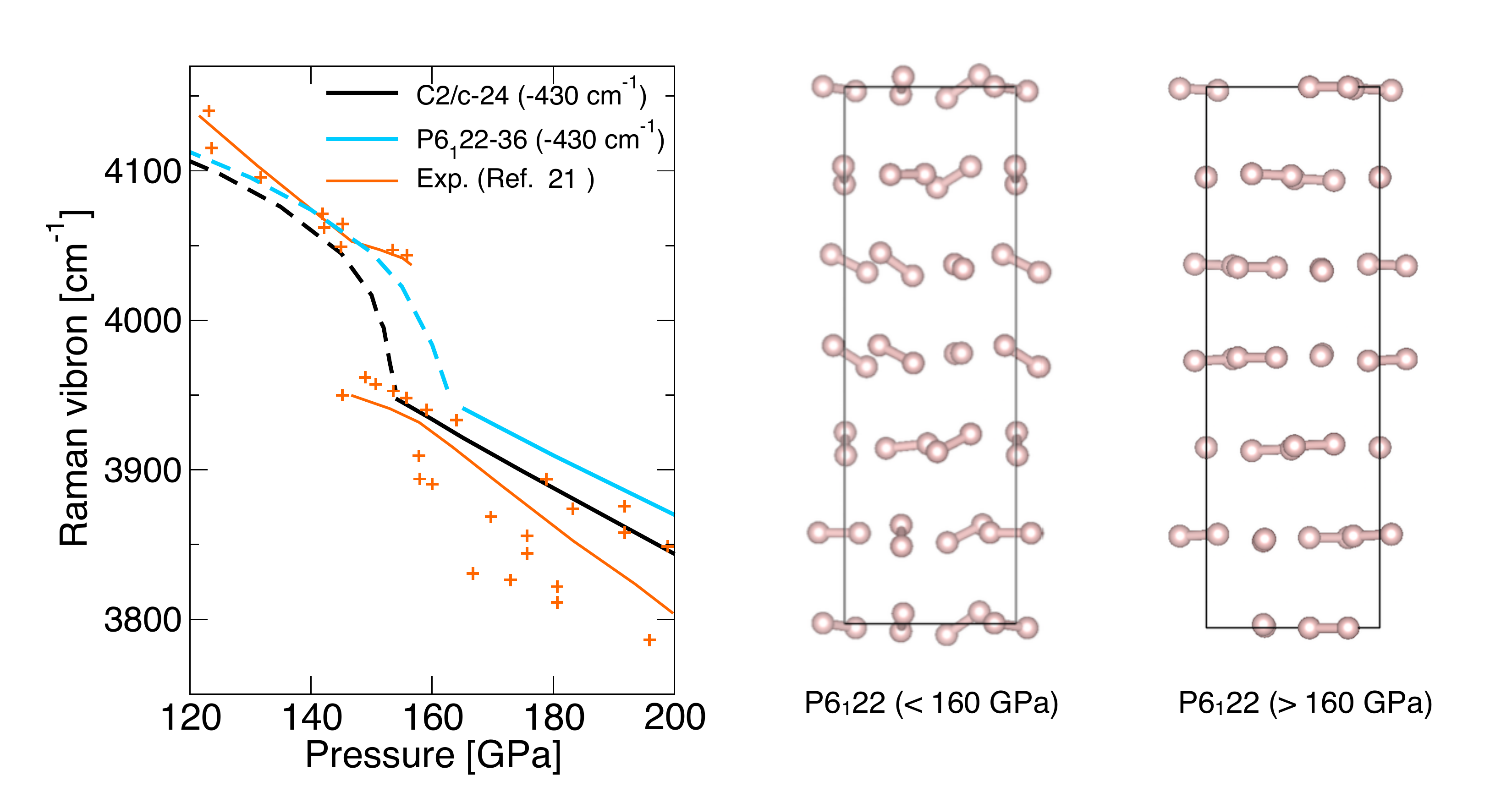}
\caption{Left: The first vibron frequency as a function of pressure
  for the C2/c-24 and P6$_1$22-36 structures. A symmetry lowering
  occurs at 155 GPa and 160 GPa respectively, marked by dashed
  lines. The theoretical results are shifted by -430 cm$^{-1}$ (a
  value close to the expected anharmonic shift) in order to compare
  the frequency shift at the transition with experiment (here taken
  from Ref.~\cite{PhysRevLett.63.2080}). Right: Structure of
  P6$_1$22-36 along the c-axis below and above the transition
  (visualized at 140 GPa).} 
\label{fig:p6122_transition}
\end{figure}

\newpage

\section{Nuclear quantum effects across the II-III phase boundary}

In order to estimate the NQEs across the II-III phase boundary, we modeled the total energy $E$ of the system (unit cell) as a function of the scalar order parameter $h$ by using a Landau-like potential whose form is
\begin{equation}
    E(h) = a + b h^2 + c h^4.
    \label{landau}
\end{equation}
We take the order parameter $h$ as the distance between the hydrogen atoms belonging to the canted (out-of-plane) H$_2$ molecules and the hcp layer of the H$_2$ centers of mass. While $a$ is an irrelevant energy shift, $b$ and $c$ are determined by the equilibrium value $h_\textrm{min}$ of the order parameter, i.e. the canted equilibrium geometry, and by the energy gain $\Delta$ of the P-1-24 phase with respect to the C2/c symmetry of phase III. Straightforwardly, $b=-2 \Delta / h_\textrm{min}^2$ and $c=\Delta / h_\textrm{min}^4$. Thus, both $b$ and $c$ are pressure ($p$) dependent through the $p$ dependence of $h$ and $\Delta$. The symmetry of the potential in Eq.~\ref{landau} is dictated by the fact that the P-1-24 crystalline symmetry allows for two degenerate equilibrium geometries (given by $h_\textrm{min}$ and $-h_\textrm{min}$), and the broken order parameter $h_\textrm{min}$ continuously goes to zero as $p \rightarrow p_c$, with $p_c$ the critical pressure of the II-III phase transition.

To include NQEs, we solve the one-dimensional Schr\"odinger equation
\begin{equation}
    \left(-\frac{1}{2 M} \nabla^2_h + E(h)\right)\Psi = E \Psi,
    \label{schroedinger}
\end{equation}
where $M$ is the effective mass given by $M=0.5 N_\textrm{mol} m_\textrm{H/D}$, with $N_\textrm{mol}$ the number of canted molecules in the unit cell, $m_\textrm{H/D}$ the proton/deuterium mass in Hartree units, and $0.5$ the reduced mass of a single H$_2$ molecule. If the resulting ground state (GS) wave function $\Psi=\Psi(h)$ of Eq.~\ref{schroedinger} is bimodal, then the system is in the broken phase II. If $\Psi$ has a peak at $h=0$, namely if the average molecular distribution is an in-plane configuration, the system is in phase III.

 \begin{figure}[b]
\includegraphics[width=0.32\textwidth]{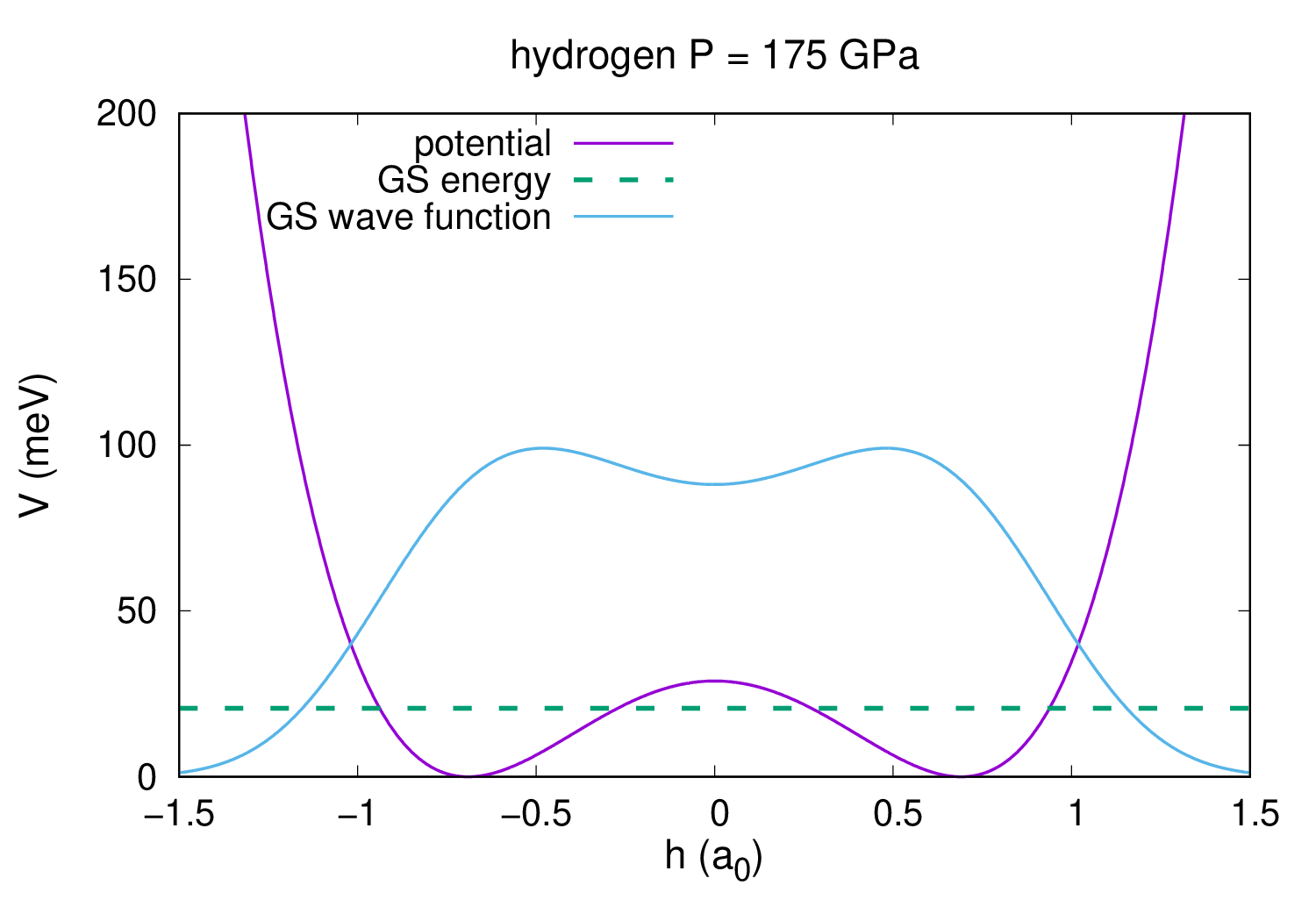}
\includegraphics[width=0.32\textwidth]{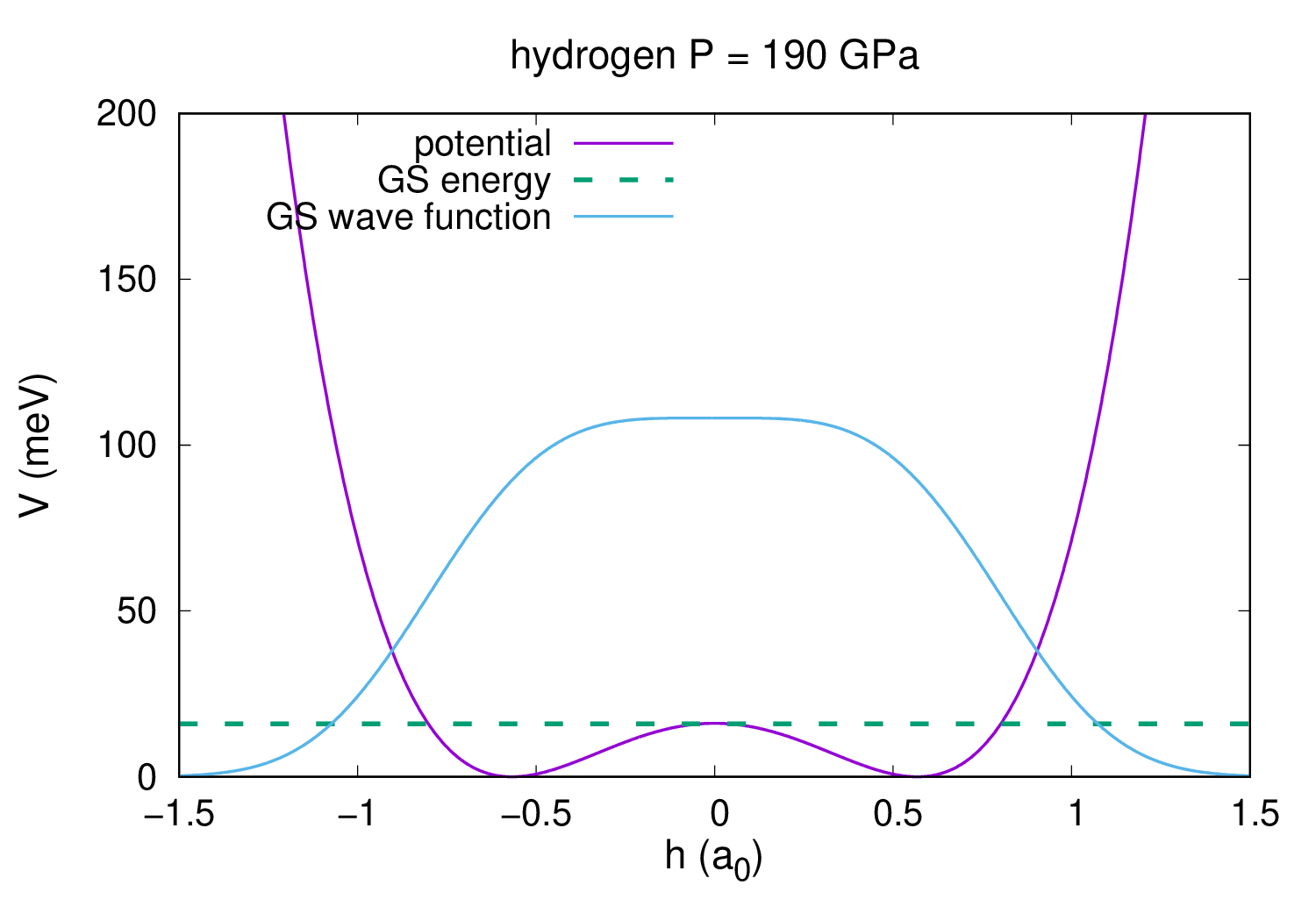}
\includegraphics[width=0.32\textwidth]{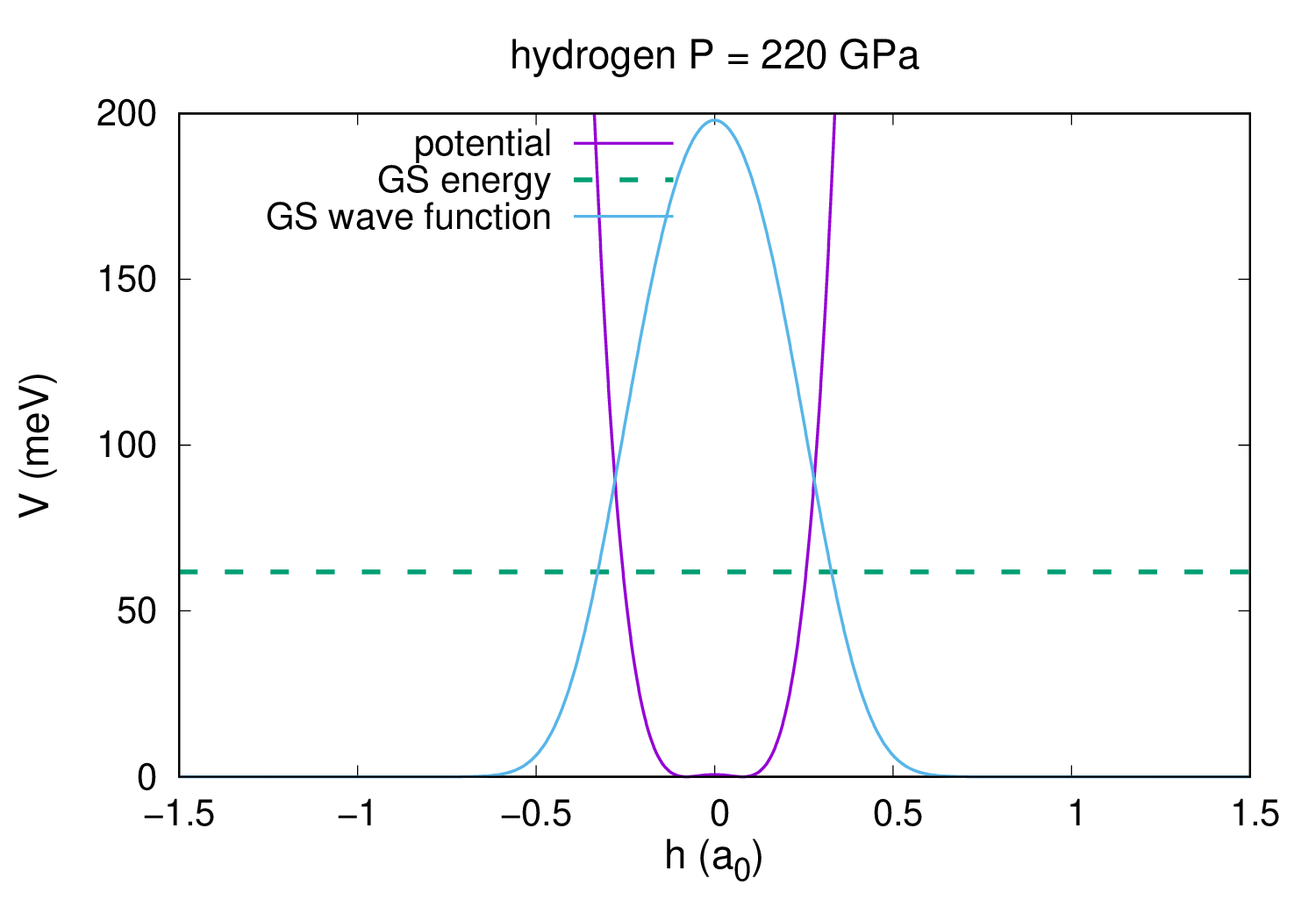}
\includegraphics[width=0.32\textwidth]{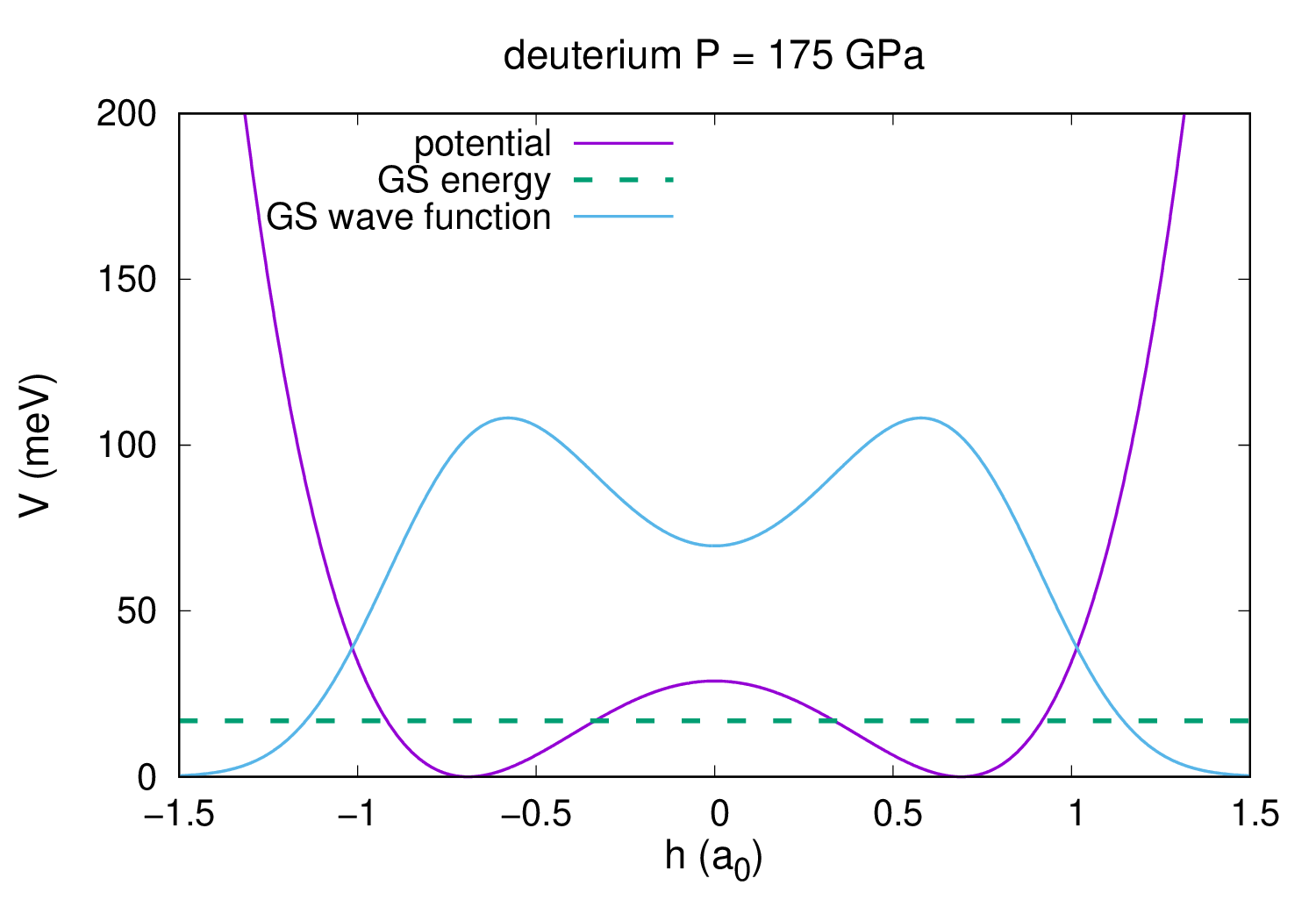}
\includegraphics[width=0.32\textwidth]{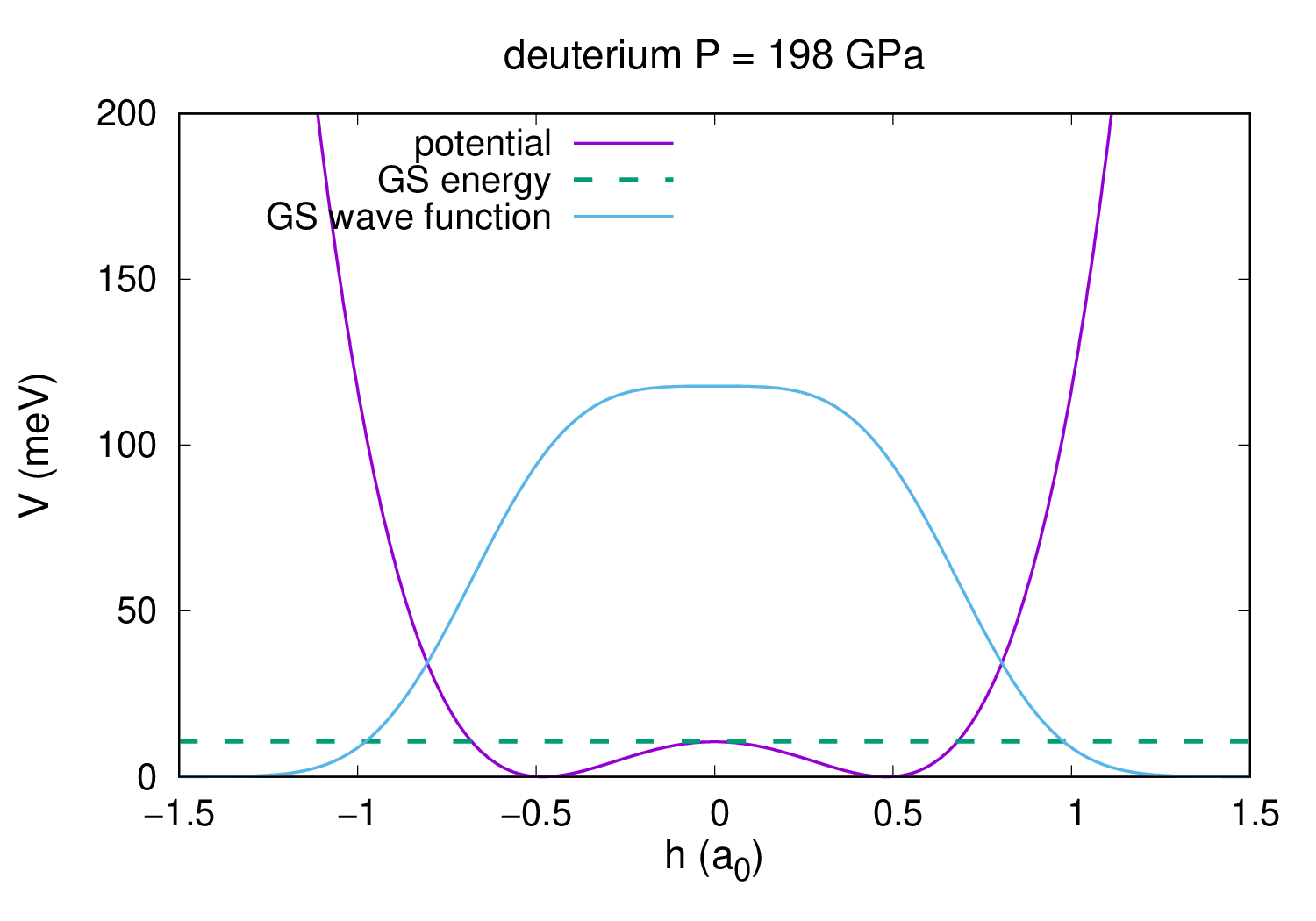}
\includegraphics[width=0.32\textwidth]{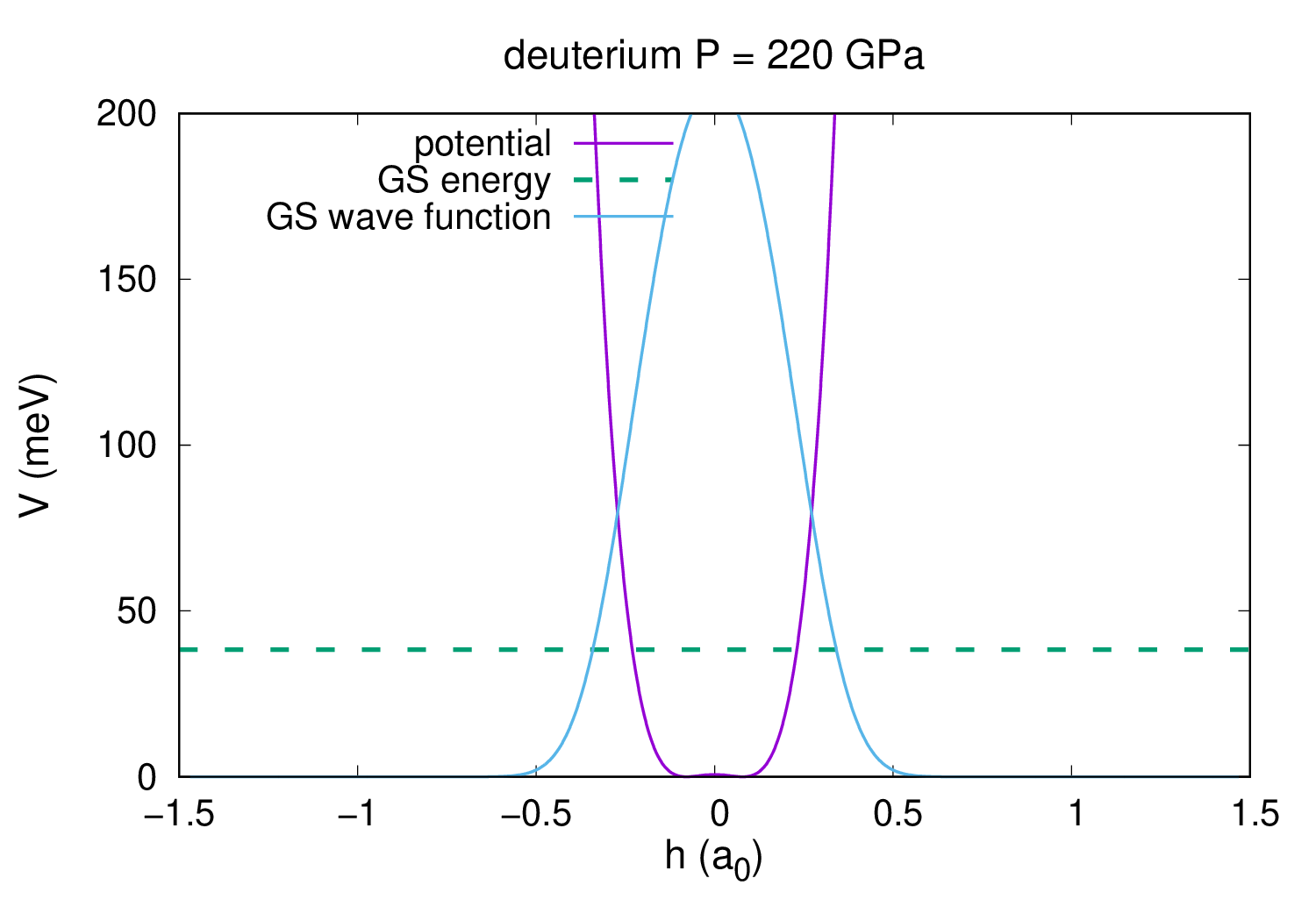}
\label{fig:nqes}
\caption{Potential (violet line), GS wave function (cyan line) and GS energy (dashed green line) for the hydrogen (upper row) and deuterium (lower row) librational degrees of freedom, as a function of the order parameter $h$, for a system 
with $N_\textrm{mol}=4$ canted dimers per unit cell of 12 molecules. The central column is computed at the transition pressure $p_c$ separating phase II and III, while the left (right)-hand side column is for $p=175$ GPa ($p=220$ GPa), deep in phase II (phase III).}
\end{figure}

The results reported in 
Fig.~\ref{fig:nqes}. 
for both hydrogen (upper row) and deuterium (lower row), are
obtained using a potential derived from \emph{ab initio} HYB$_\textrm{opt}$ geometries and energies. One can notice that, according to the model in Eq.~\ref{schroedinger}, the transition (shown in the central panels) takes place at $p_c \in [190-198]$ GPa ($p_c \in [198-204]$ GPa) for hydrogen (deuterium), at variance with the ``static lattice'' system, where the transition happens at $\approx$ 220 GPa, as yielded by our advanced calculations (HYB$_{\rm opt}$, many-body RPA and QMC, see main text). The pressure range depends on the number $N_\textrm{mol}$ of canted molecules taken into account in the model. Indeed, as shown in Figs.~6 and 7, not all out-of-plane molecules share the same $h$ in the 12-H$_2$ unit cell. It is possible to distinguish three groups, made of 4 molecules each, one with the largest $h$, the second with an intermediate value, the last one with the smallest out-of-plane component. Solving Eq.~\ref{schroedinger} with $N_\textrm{mol}=4$ and $N_\textrm{mol}=8$ yields an estimate of the $p_c$ range. Two conclusions can be drawn from this analysis:
\begin{itemize}
    \item the estimated isotopic effect across the II-III phase boundary amounts approximately to 10 GPa;
    \item in hydrogen, the NQEs shift due to the librational degrees of freedom is $\approx$ 25 GPa from the static lattice pressure toward lower values.
\end{itemize}

\section{Bader analysis and infrared spectra}
Below we present the results from the Bader analyses of C2/c-24 and
P-1-24. The structures are visualized from two perspectives in
Fig.~\ref{fig:structures}. In Fig.~\ref{fig:polarization} we plot the polarization as function pressure for
the six inequivalent molecules in P-1-24. After the transition there
are three inequivalent molecules that lie in plane and the symmetry
increases to C2/c. At the same time the polarization increases. A
similar analysis was performed in Ref.~\cite{doi:10.1063/1.3679749}
for C2/c-24 and P6$_3$/m-16.  

In Fig.~\ref{fig:IR} we plot the infrared (IR) intensity of the IR active vibron
modes at different pressures. Comparing to the experimental results of
Ref. \cite{PhysRevLett.70.3760} we see a very similar behavior.

\begin{figure}[ht!]
\includegraphics[scale=0.7]{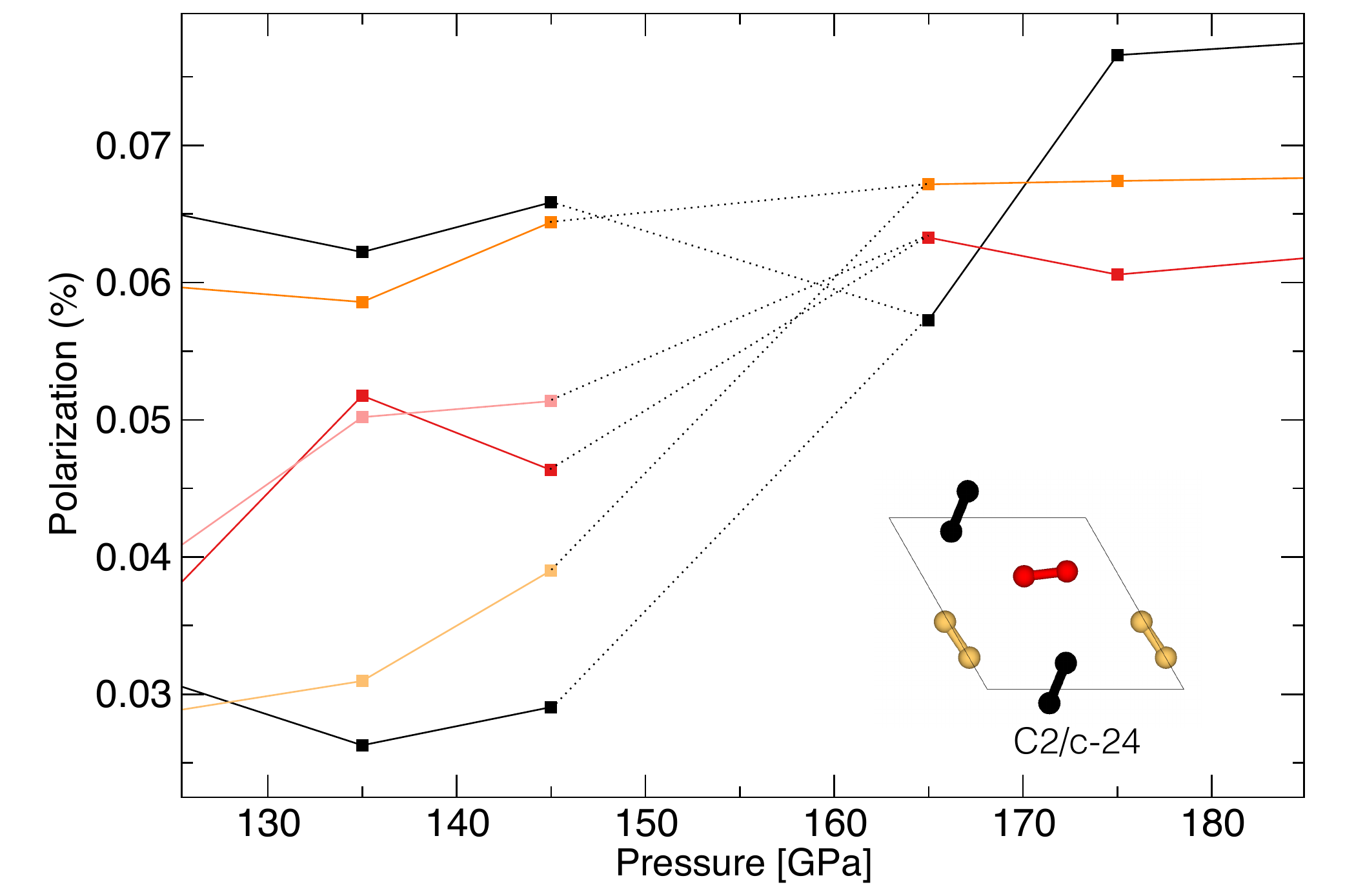}
\caption{The polarization of the molecular units in C2/c-24 and P-1-24
  obtained with a Bader analysis. While there are three inequivalent
  molecules in C2/c-24 there are six in P-1-24. The inset shows a
  single layer of C2/c-24 where the molecules are colored by their
  polarization in the main figure.}
\label{fig:polarization}
\end{figure}

\begin{figure}[h!]
\includegraphics[scale=0.7]{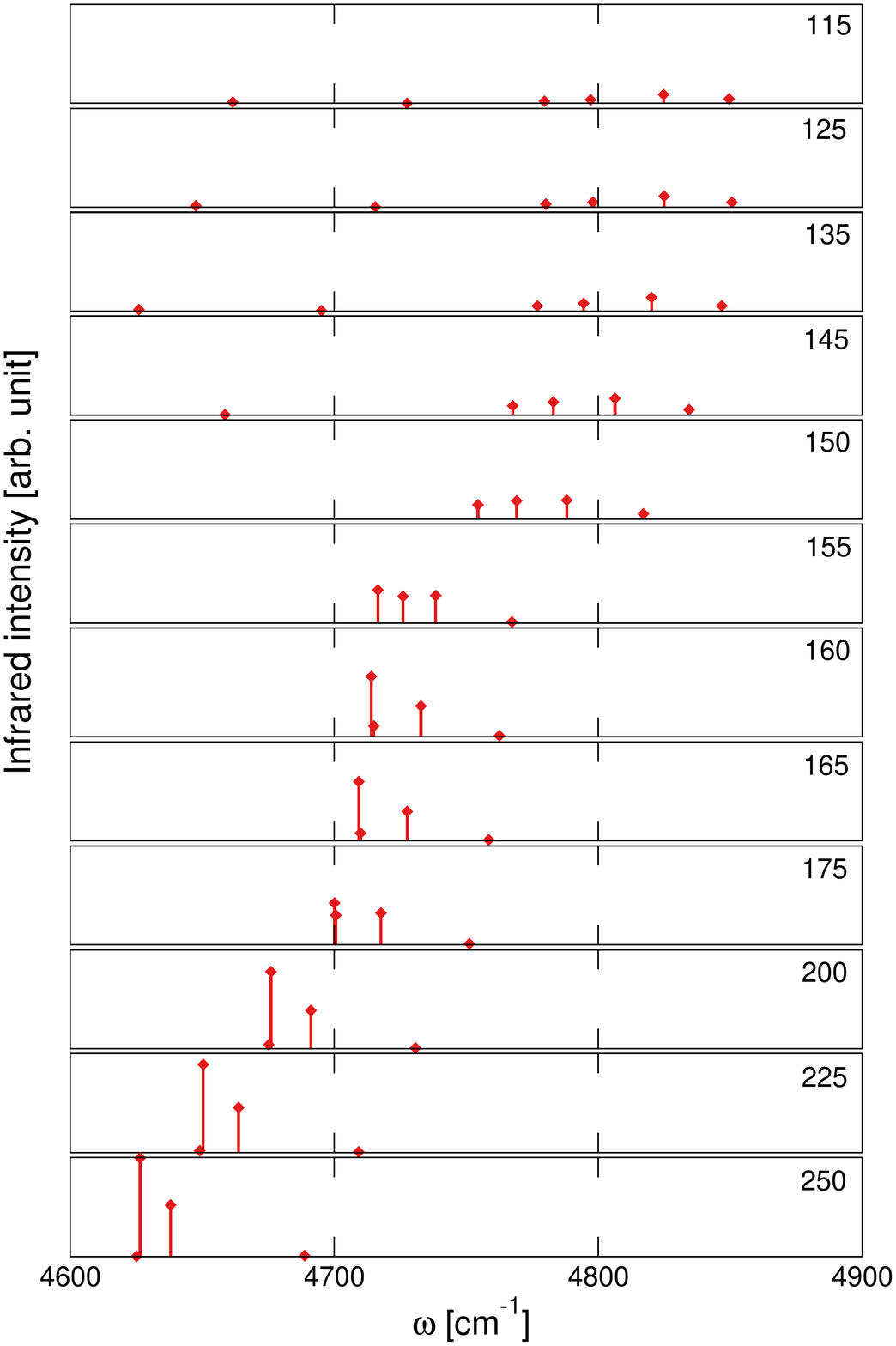}
\caption{The intensity of the infrared active vibrons as a function of
  pressure for C2/c-24. At approximately 155 GPa the C2/c-24 phase
  turns into P-1-24, signaled by the frequency shift and increase in
  intensity.}
\label{fig:IR}
\end{figure}

\end{widetext}
\end{document}